\documentclass[aps,prb,reprint,superscriptaddress,longbibliography]{revtex4-2}
\usepackage{graphicx}
\usepackage{textcomp}
\usepackage{gensymb}
\usepackage{amsmath, amssymb}
\usepackage[colorlinks = true, citecolor = magenta]{hyperref}
\usepackage{braket}
\usepackage{natbib}
\usepackage[utf8]{inputenc}
\usepackage{float}
\usepackage{graphicx}
\usepackage{overpic}
\usepackage[normalem]{ulem}
\usepackage{xcolor}
\usepackage{hyperref}
\begin{document}

\title{Post-Selection-Based Stochastic Quantum Battery}

\author{Aman~Verma}
\email{v.aman@iitg.ac.in}

\author{Roson~Nongthombam}
\email{n.roson@iitg.ac.in}

\author{Amarendra~K.~Sarma}
\email{aksarma@iitg.ac.in}

\affiliation{Department of Physics, Indian, Institute of Technology Guwahati, Guwahati-781039, (India)}

\date{\today}

\begin{abstract}
Quantum batteries are quantum systems that store useful energy, which can be subsequently extracted to perform work. In this paper, we investigate the charging dynamics of the quantum battery under continuous measurements and post-selection. By post-selecting the trajectories with no quantum jump $\ket{1}\rightarrow\ket{0}$ in a three-level system, the dynamics is effectively confined to a two-level $(\ket{2}$-$\ket{1})$ manifold. We analyze the resulting charging dynamics under both $\sigma_x$- and $\sigma_y$-driving protocols and investigate the roles of detuning, coherent driving, non-linearity, and measurement backaction. The generation and dynamics of quantum coherence, which plays a pivotal role in quantum battery, are also analyzed under post-selection. The advantage and comparison of our results over those of a generic two-level system and with the Lindblad dynamics are also highlighted. Our work explores the interplay between the measurement axis and the driving protocols that govern the charging rate, ergotropy, and quantum coherence of the non-Hermitian, highly tuneable, and controllable quantum battery.
\end{abstract}

\maketitle
\section{Introduction}
Non-Hermitian quantum systems realized by post-selecting quantum trajectories in which no quantum jumps occur have been experimentally demonstrated in a three-level superconducting circuit QED system with energy levels $\ket{2}$, $\ket{1}$, and $\ket{0}$ \cite{PhysRevLett.128.160401,Naghiloo2019}. By post-selecting trajectories with no $\ket{1}\rightarrow\ket{0}$ quantum jumps, the effective ensemble dynamics remains confined to the $\ket{2}$--$\ket{1}$ subspace, thereby realizing an effective postselected non-Hermitian two-level system. A distinctive feature of non-Hermitian systems is the existence of exceptional points (EPs), where both the eigenvalues and the corresponding eigenvectors coalesce. During the past two decades, exceptional points have attracted considerable interest due to their remarkable physical properties and potential applications, including enhanced sensing \cite{kmtx-7x9d}, control of open quantum systems \cite{DEY2019125931}, entanglement protection \cite{PhysRevA.100.063846, PhysRevLett.131.100202}, and applications in photonic \cite{Miri_2019} and other classical systems \cite{Peng2014, feng2017, hodaei2014, PhysRevLett.115.040402, xiao2017, peng}. Among non-Hermitian systems, an important class is formed by parity-time ($\mathcal{PT}$)-symmetric systems \cite{doi:10.1142/q0178, PhysRevLett.80.5243, doi:10.1142/S0219887810004816, Ashida02072020}. These systems exhibit a characteristic $\mathcal{PT}$-symmetry-breaking transition, in which the spectrum changes from being entirely real to consisting of complex-conjugate eigenvalue pairs as a system parameter is varied. This phase transition occurs at an exceptional point and is accompanied by a qualitative change in the dynamics of the system, from an oscillatory (unbroken $\mathcal{PT}$) regime to an overdamped (broken $\mathcal{PT}$) regime\cite{SciPostPhys.9.4.052} and various other applications \cite{hodaei2017, xu2016, zhang2017, PhysRevB.100.134505, shi2016, chen2017, lau2018}. 

More recently, exceptional points of the Liouvillian superoperator, often referred to as genuine quantum exceptional points, have attracted significant attention owing to their ability to influence and control the dynamics of open quantum  \cite{Lin2025, rlhg-8fn7}. In the three-level system considered here, post-selection of trajectories with no $\ket{2}\rightarrow\ket{1}$ and $\ket{1}\rightarrow\ket{0}$ quantum jumps leads to an effective non-Hermitian evolution whose Liouvillian spectrum can exhibit a third-order exceptional point, where three eigenvalues and their corresponding eigenmodes simultaneously. Interestingly, this higher-order degeneracy is highly sensitive to the underlying measurement record. In particular, the inclusion of trajectories containing $\ket{2}\rightarrow\ket{1}$ quantum jumps lifts the third-order degeneracy and modifies the Liouvillian spectrum. As a consequence, the higher-order exceptional point splits into lower-order degeneracies, resulting in qualitatively different dynamical behavior \cite{role_of_inefficient, homodyne_roson}. 

On the other hand, quantum batteries are quantum thermodynamic devices designed to store and deliver useful energy in the form of extractable work, enabling the implementation of various quantum tasks \cite{PhysRevE.87.042123, RevModPhys.96.031001, Binder_2015, PhysRevLett.118.150601, PhysRevLett.111.240401, PhysRevLett.120.117702}. In recent years, quantum batteries have been extensively investigated in a wide range of physical platforms, including quantum dots\cite {Loukhssami_2026}, optomechanical systems \cite{PhysRevA.110.062204, 9vv8-s8r1}, superconducting circuits \cite{PhysRevA.107.023725}, spin \cite{xqtv-qbyk, Evangelakos2025, Sun_2025}, cavity-QED systems \cite{Wang_2026}, topological photonic waveguide \cite{PhysRevLett.134.180401}, and various many-body systems \cite{Wang_2026}, also with non-Hermitian charging \cite{PhysRevA.109.042207}. Furthermore, the role of quantum entanglement and coherence has also been an important topic in the study of quantum batteries, as both resources can significantly influence the charging dynamics, energy-storage capacity, and extractable work of quantum battery systems \cite{10.1116/5.0184903, PhysRevLett.129.130602}. More recently, continuously monitored stochastic quantum batteries have emerged as a promising paradigm, where feedback control can be used to enhance the charging process and improve the extractable work \cite{Mitchison2021chargingquantum, f591-hx6x}. 

In this work, we investigate the charging dynamics of a post-selection-based non-Hermitian quantum battery. We highlight the role of nonlinear normalization and trajectory fluctuations arising from continuous measurement and post-selection, and demonstrate how these effects lead to a distinct charging behavior and an advantage over a generic two-level non-Hermitian system.

The remainder of this paper is organized as follows. In Sec.~\ref{sec:hybrid_measurement}, we introduce the three-level hybrid continuous-measurement model, describe the post-selection scheme, and derive the effective non-Hermitian dynamics of the corresponding two-level system. In Sec.~\ref{sec:energitics}, we investigate the charging dynamics and discuss various characterizers of the quantum battery. In Sec.~\ref{sec:results} we discuss the post-selection-based non-Hermitian advantage by comparing our results with that of a generic non-Hermitian two-level system and highlight our main results. Finally, in Sec.~\ref{sec:conclusion}, we conclude.

\begin{figure}
    \centering
    \includegraphics[width=1\linewidth]{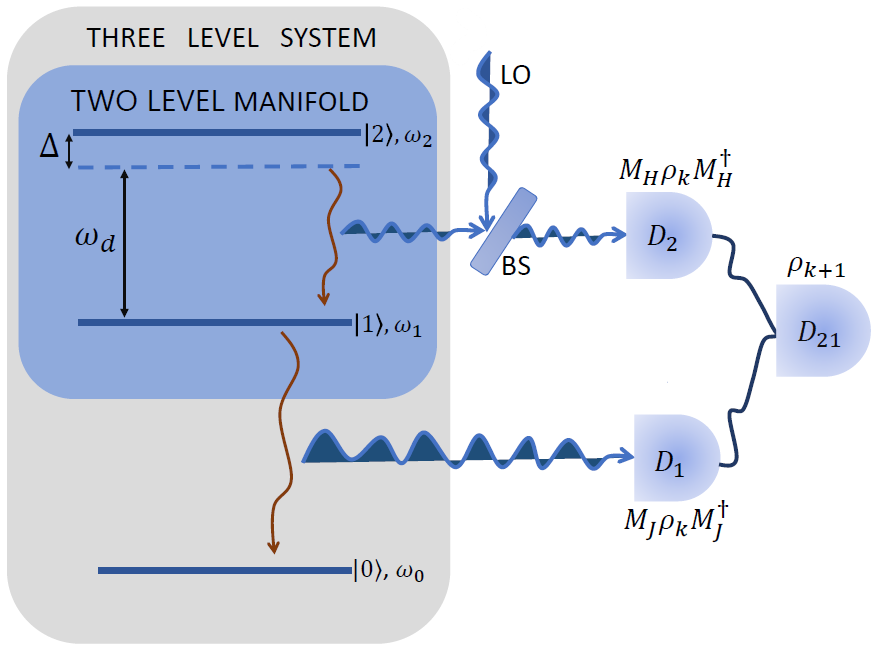}
    \caption{Schematic of a three-level system as quantum battery with hybrid continuous quantum measurement. $\omega_d$ represents the frequency of the coherent drive applied between $\ket{2}$ and $\ket{1}$ states. The second and first excited states decay via $\ket{2}\rightarrow\ket{1}$ and $\ket{1}\rightarrow \ket{0}$ transitions with decay rates $\Gamma_e$ and $\Gamma_g$, respectively. Both the decays are monitored via measurement setup contain detectors $D_2,~D_1~\text{and}~D_{21}$. In the $\ket{2}\rightarrow \ket{1}$ transition the state is updated via the Kraus matrix corresponding to homodyne measurement $M_H$ whereas in the case of $\ket{1}\rightarrow \ket{0}$ decay state is updated via jump Kraus matrix $M_J$. $LO$ represents the local oscillator with frequency $\omega_{LO} = \omega_2-\omega_1$ with phase $\theta $. $BS$ represents the beam a $50-50$ splitter. The detector $D_{21}$ gives the final state of the system along with the measurement record.}
    \label{fig: model}
\end{figure}

\section{Hybrid Measurement of a Three-Level System and Post-selected effective two-level manifold} 
\label{sec:hybrid_measurement}

Consider a three-level system with energy eigenstates $\ket{2}$, $\ket{1}$, and $\ket{0}$, corresponding respectively to the second excited state, first excited state, and ground state (See Fig.~\ref{fig: model}). The system allows two transitions: $\ket{2}\rightarrow\ket{1}$ and $\ket{1}\rightarrow\ket{0}$, with decay rates $\Gamma_e$ and $\Gamma_g$, respectively. We consider a hybrid measurement scheme similar to that discussed in Ref.~\cite{homodyne_roson}. In this setup, the transition $\ket{2}\rightarrow\ket{1}$ is continuously monitored via a weak homodyne measurement, while the transition $\ket{1}\rightarrow\ket{0}$ is monitored through photo-detection, corresponding to a projective jump-type measurement. The homodyne measurement can be performed by inserting a beam splitter with a local oscillator between the signal and the detector. The Kraus matrix for the homodyne measurement is given as follows \cite{lewalle2020}  
\begin{equation}
    M_H = \sqrt{N}e^{-r^2dt/4}\begin{bmatrix}
    \sqrt{1-\Gamma_edt} & 0 & 0 \\
    r~dt~ e^{-i\theta}\sqrt{\Gamma_e} & \sqrt{1-\Gamma_gdt} & 0 \\
    0 & \sqrt{\Gamma_gdt~} \hat{a}^\dagger_g & 1
    \end{bmatrix},
\end{equation}
where $r$ is the measurement record for the homodyne signal and $\theta$ is the phase difference between signal $\ket{2}\rightarrow \ket{1}$ transition and the local oscillator. Whereas the Kraus matrix for the jump $\ket{1}\rightarrow\ket{0}$ decay is given as 
\begin{equation}
    M_J = \sqrt{N}e^{-r^2dt/4}\begin{bmatrix}
    0 & 0 & 0 \\
    0 & 0 & 0 \\
    0 & \sqrt{\Gamma_gdt~}  & 0
    \end{bmatrix},
\end{equation}
The Bayesian state update equation is then used to update the state of the system conditioned on the measurement signal and/or the measurement type at time $dt$ as follows \cite{PhysRevA.92.032125}
\begin{align}
    \rho_c(t+dt) = \frac{UM_{H,J}\rho_c(t) M^\dagger_{H,J}U^\dagger }{\text{Tr}\left(UM_{H,J}\rho_c(t) M^\dagger_{H,J}U^\dagger \right)}
    \label{eq: state_update_equation}
\end{align}
Here, the denominator represents the probability of obtaining a particular measurement outcome conditioned on the state $\rho_c(t)$, while $U=e^{-iHdt}$ denotes the unitary time-evolution operator. We consider a coherent drive between the states $\ket{2}$ and $\ket{1}$. The corresponding Hamiltonian in the rotating frame of external drive is given as (See Appendix \ref{app:ergo_hamiltonian_derivation} for details)

\begin{align}
H = \Delta \ket{2}\bra{2} +  \Omega \left( \ket{2}\bra{1} + \ket{1}\bra{2} \right),
\label{eq: Hamiltonian}
\end{align}

where $\Delta = (\omega_2 - \omega_1)-\omega_d$ is the detuning between the system and the external drive (considering $\omega_0=0$) and $\Omega$ is the drive strength. Under this coherent ($\sigma_x$) drive, the system undergoes coherent oscillations between the states $\ket{2}$ and $\ket{1}$. The oscillation continues until a quantum jump associated with the $\ket{1}\rightarrow\ket{0}$ transition occurs, after which the system decays to the ground state $\ket{0}$ and remains there for the rest of the evolution. By taking the average of such many trajectories, we get the ensemble-averaged dynamics, which corresponds to a mixed state equivalently described by the Lindblad master equation \cite{10.1093/acprof:oso/9780199213900.001.0001}. The Stochastic master equation for such hybrid measurement model is given by \cite{homodyne_roson}

\begin{multline}
    d\rho_c = -i[H,\rho_c] + \Gamma_e \mathcal{D}[\ket{1}\bra{2}]dt - \frac{\Gamma_g}{2}\{\ket{1}\bra{1}, \rho_c\}dt
    \\
    +\sqrt{\Gamma_g}\mathcal{H}[\ket{1}\bra{2}e^{i\theta}]\rho_c dW + \Gamma_g \langle\ket{1}\bra{1} \rangle\rho_c dt
    \\    
    + \left( \frac{\ket{0}\bra{1}\rho_c \ket{1}\bra{0}}{\langle \ket{1}\bra{1}\rangle} - \rho_c\right)dN 
    \label{eq: SME}
\end{multline}

where $dN$ is the number of counts for $\ket{1}\rightarrow \ket{0}$ jump process in time interval $[t, t+dt]$ with the mean value $\langle dN\rangle=\Gamma_g \langle \ket{1}\bra{1}\rangle dt$, $H$ is the Hamiltonian of the system,  $\mathcal{D}[\ket{1}\bra{2}]\rho_c$ is the Lindblad dissipator terms given as 
\begin{align}
    \mathcal{D}[L]\rho = L\rho L^\dagger - \frac{1}{2}\left\{ L^\dagger L, \rho\right\}
    \label{eq:lindblad_dissipater}
\end{align}
with $L = \sqrt{\Gamma_e}\ket{1}\bra{2}$ being the Lindblad jump operator. The term with $dW$ in Eq.~(\ref{eq: SME}) is known as $inovation~term$ and the superoperator $\mathcal{H}$ is given as   

\begin{multline}            \mathcal{H}\left[\ket{1}\bra{2}e^{i\theta}\right]\rho_c = e^{i\theta} \ket{1}\bra{2}\rho_c + e^{-i\theta}\rho\ket{2}\bra{1}  
    \\ -\langle e^{i\theta}\ket{e}\bra{2} + e^{-i\theta}\ket{2}\bra{1}\rangle\rho_c
    \label{eq:inovation_terms}
\end{multline}
and $dW$ is the Gaussian Wiener increment with zero mean and typical realizations of the $dW$ scale as $\sqrt{dt}$. Lastly, $\Gamma_g \langle \ket{1}\bra{1}\rangle \rho_cdt$ is a non-linear term that accounts for normalization.   

\begin{figure}[t]
    \centering
    \includegraphics[width=0.49\linewidth]{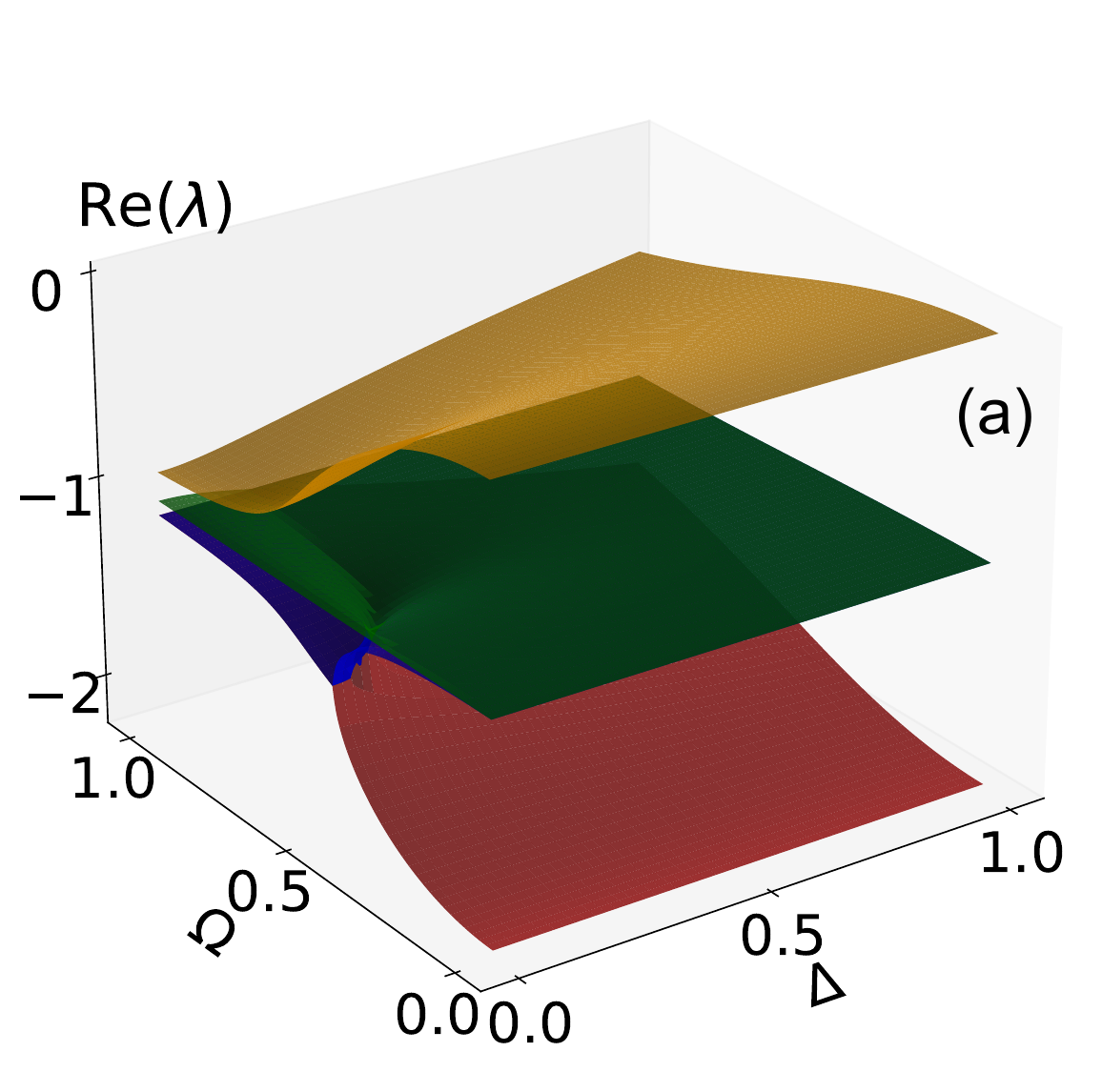} 
    \includegraphics[width=0.49\linewidth]{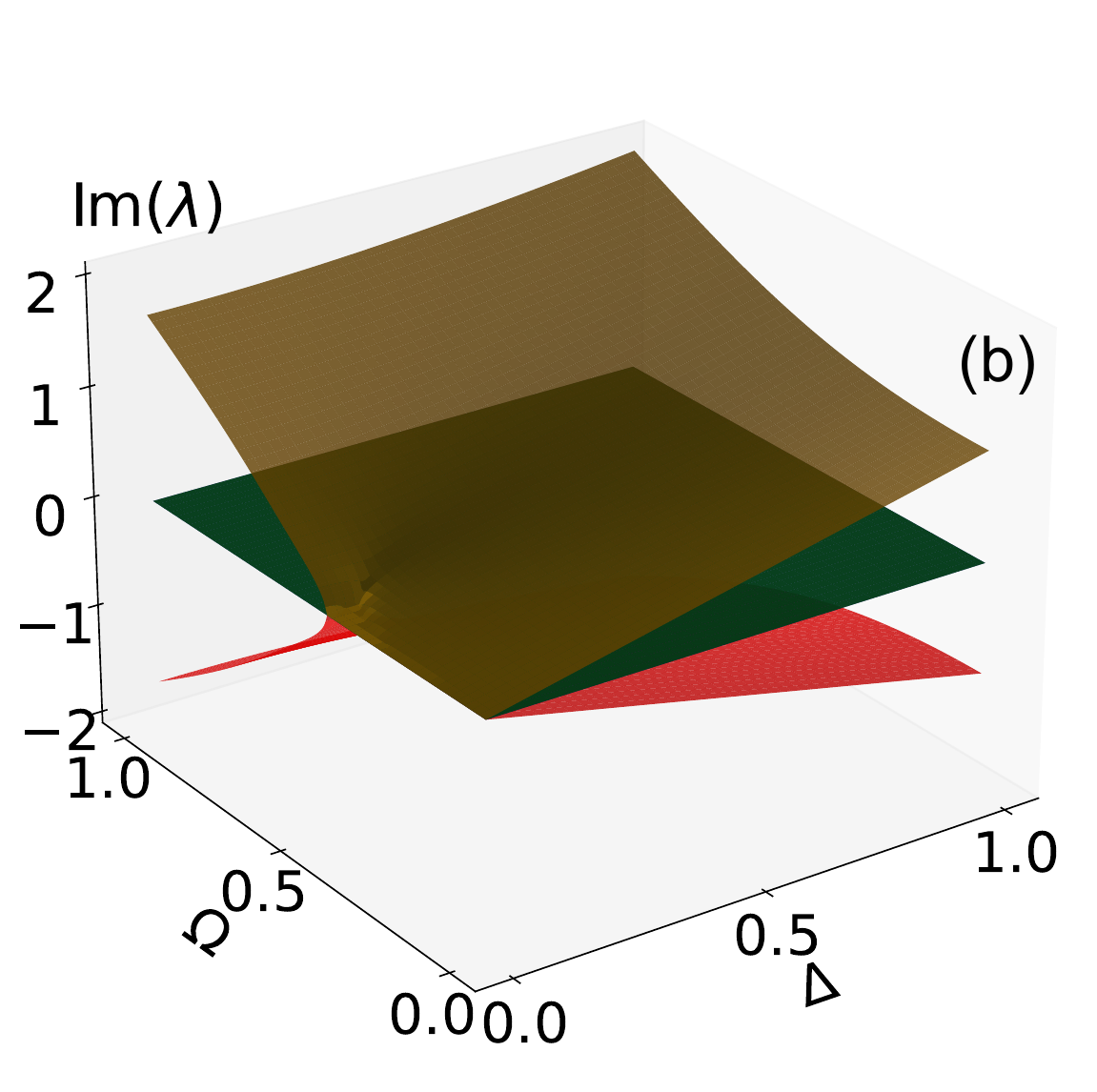}%
    \caption{(a) Real and (b) imaginary 3D representation of the  eigenvalues of the Liouville superoperator given in Eq.~(\ref{eq: Liouville_super}) for $\sigma_x$-drive case. Other parameters are $\Gamma_g=2~\text{MHz}$ and $\Gamma_e=0.2~\text{MHz}$.}
    \label{fig:eigenvalue_plot}
\end{figure}

In experimental scenarios, out of all possible quantum trajectories generated by Eq.~(\ref{eq: state_update_equation}), one can post-select the trajectories for which no $\ket{1}\rightarrow \ket{0}$ transition. Since no quantum jump from $\ket{1}$ to $\ket{0}$ takes place, the system effectively remains confined within the $\ket{2}$--$\ket{1}$ manifold throughout the evolution. Setting $dN=0$, we get the mater equation for post-selected effective dynamics as 

\begin{multline}
d\rho_c
=
-i[H,\rho_c]\,dt
+\Gamma_e \mathcal{D}[\ket{1}\bra{2}]\,\rho_c\,dt  - \frac{\Gamma_g}{2}
\{\ket{1}\bra{1},\rho_c\}\,dt
\\
\sqrt{\Gamma_e}\,\mathcal{H}[\ket{1}\bra{2}e^{i\theta}]\,\rho_c\,dW  + \Gamma_g \langle \ket{1}\bra{1}\rangle \rho_cdt.
\label{eq:post_SME}
\end{multline} 

The Lindblad master equation dynamics can be obtained by taking the ensemble average of the post-selected SME (\ref{eq:post_SME}) as 

\begin{align}
\frac{d\rho(t)}{dt}
&=
-i[H,\rho(t)] +\Gamma_e\mathcal{D}[\ket{1}\bra{2}]\,\rho(t)
\nonumber\\
&\quad
-\frac{\Gamma_g}{2}\left\{\ket{1}\bra{1},\rho(t)\right\}
\nonumber\\
&\equiv \mathcal{L}[\rho(t)].
\label{eq:Lindblad_ME}
\end{align}

where $\rho(t)=\langle\rho_c(t)\rangle$ denotes the ensemble-averaged density matrix obtained by averaging over all conditioned quantum  and $\mathcal{L}$ is the Liouville operator (Appendix \ref{appendix A}). Note that the nonlinear term in Eq.~\eqref{eq:post_SME} is not included while obtaining the Liouville. The eigenvalues of this Liouville super operator are shown in Fig.~\ref{fig:eigenvalue_plot} as a function of $\Delta$ and $\Omega$. It can be observed that for $\Delta \neq0$, the degeneracy at the exceptional point is lifted and the eigenvalues repel each other, causing the scenario of lifted-EP. While the Liouvillian provides useful insight into the dynamics near the critical point, the ensemble-averaged dynamics is correctly captured by averaging the conditioned trajectories generated using Eq.~\eqref{eq: SME}, or equivalently Eq.~\eqref{eq: state_update_equation}. This approach properly accounts for the nonlinear contribution arising from the normalization associated with post-selection.

 $\textit{}{Note}$ (a): The analysis presented so far for the hybrid model with homodyne measurement can also be extended to the case of continuous quantum photodetection. In the later case one can further post-select the trajectories excluding $\ket{2}\rightarrow\ket{1}$ quantum jump. Such trajectories evolve non-unitarily with the non-Hermitian Hamiltonian $H_{eff} = H -i\frac{\Gamma_g}{2}\ket{1}\bra{1} - i\frac{\Gamma_e}{2}\ket{2}\bra{2}$. The details for such model is presented in Appendix \ref{App:jump_measurement}. 
 \\
 $\textit{Note}$ (b): We have considered the efficient measurement case for simplicity, where no photons can leak out and go undetected. We have analyzed the role of inefficient measurements for both decay channels in the case of photo-detection scheme in Ref~\cite{role_of_inefficient}. 
 
\begin{figure}[t]
    \centering
    \includegraphics[width=0.49\linewidth]{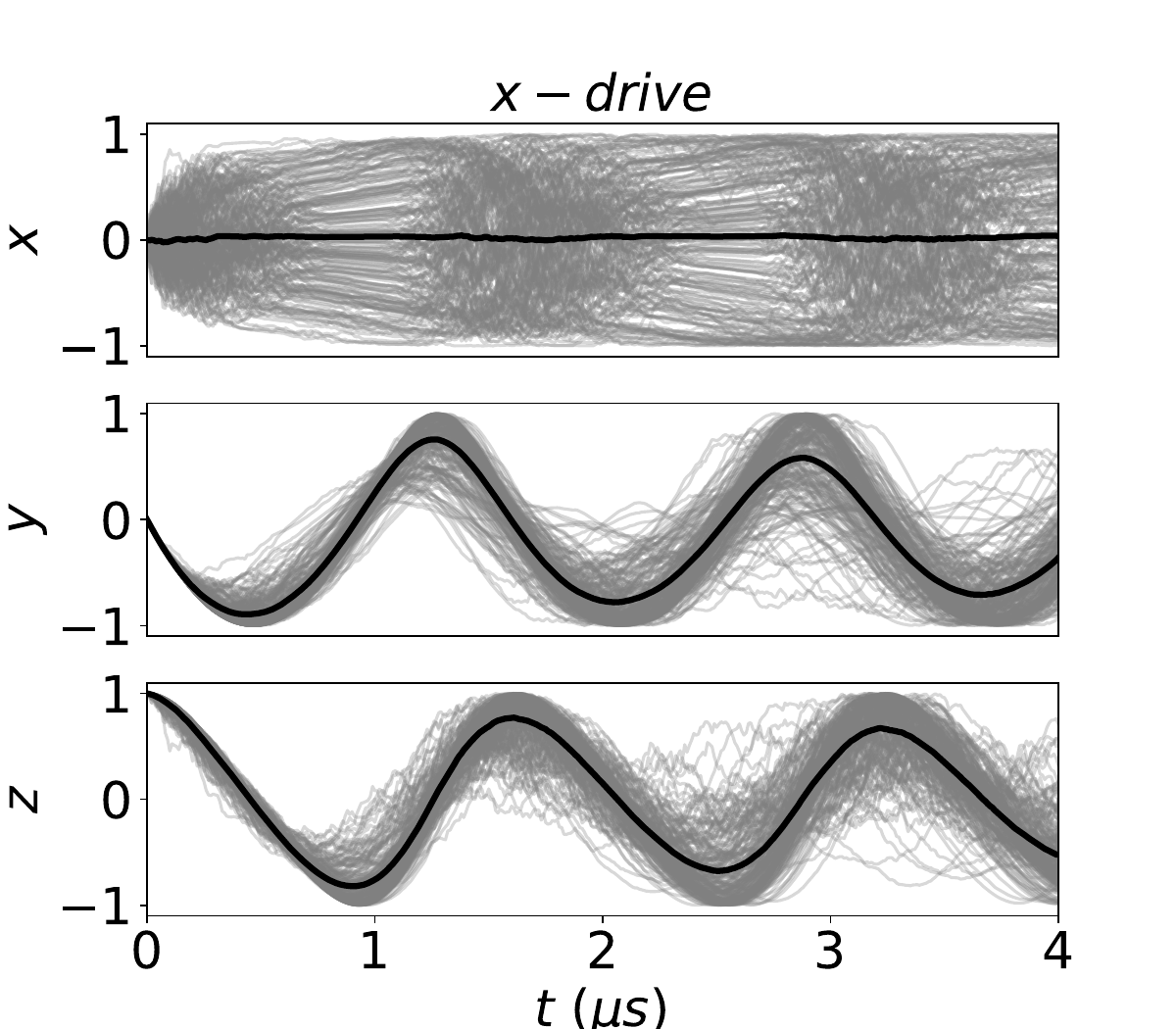}
    \includegraphics[width=0.49\linewidth]{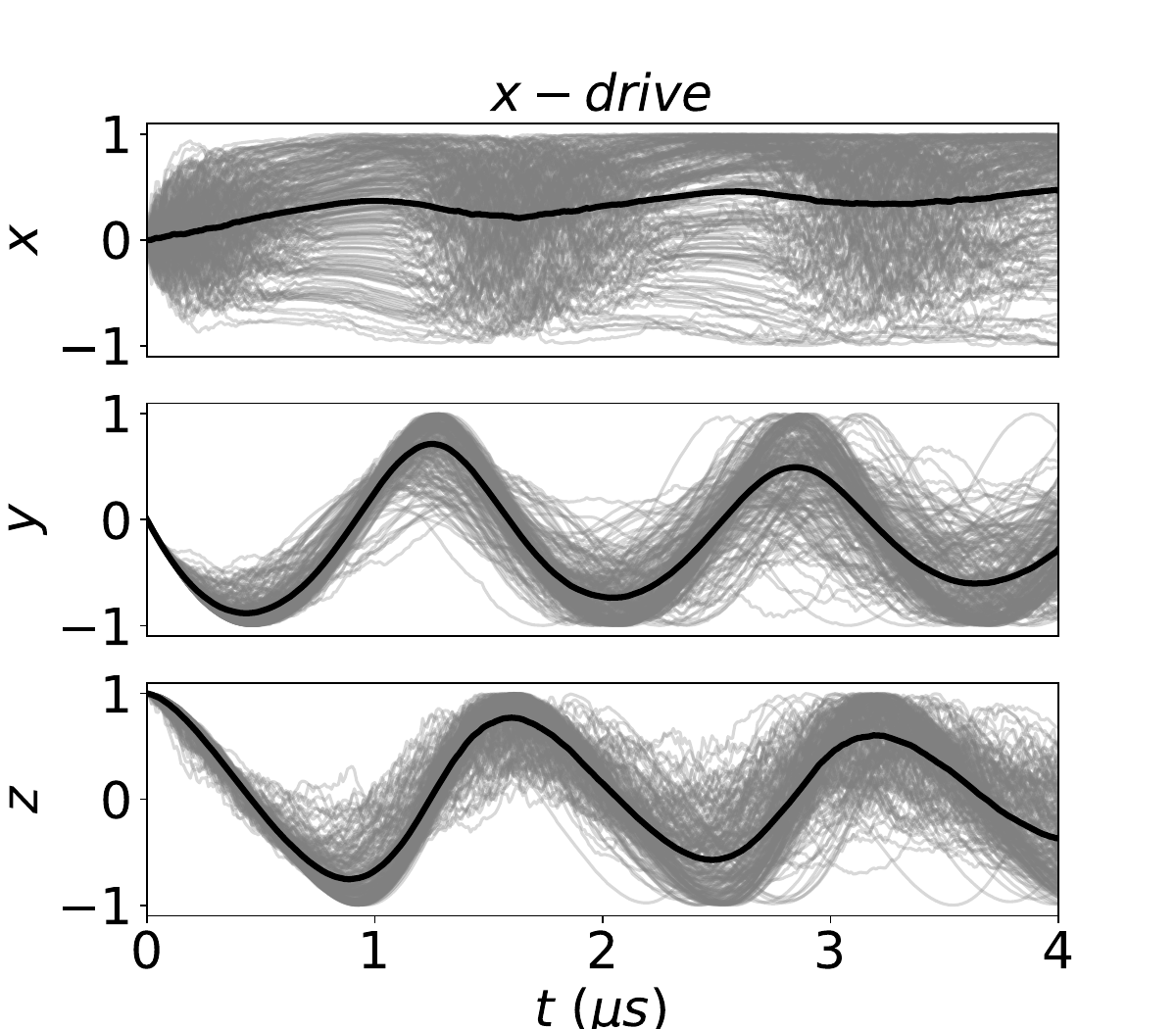}
    \caption{Bloch components trajectory evolution for $\sigma_x$-measurement with $\sigma_x$-drive. Black solid lines represents their respective ensemble average. Left and right column corresponds to $\Delta=0$- and $\Delta=0.5~\text{MHz}$, respectively. Other parameters are $\Gamma_g=2~\text{MHz}$, $\Gamma_e=0.2~\text{MHz}$ and $\Omega = 2~\text{MHz}$.}
    \label{fig:bloch_traj_x}
\end{figure}

\begin{figure}[t]
    \centering
    \includegraphics[width=0.49\linewidth]{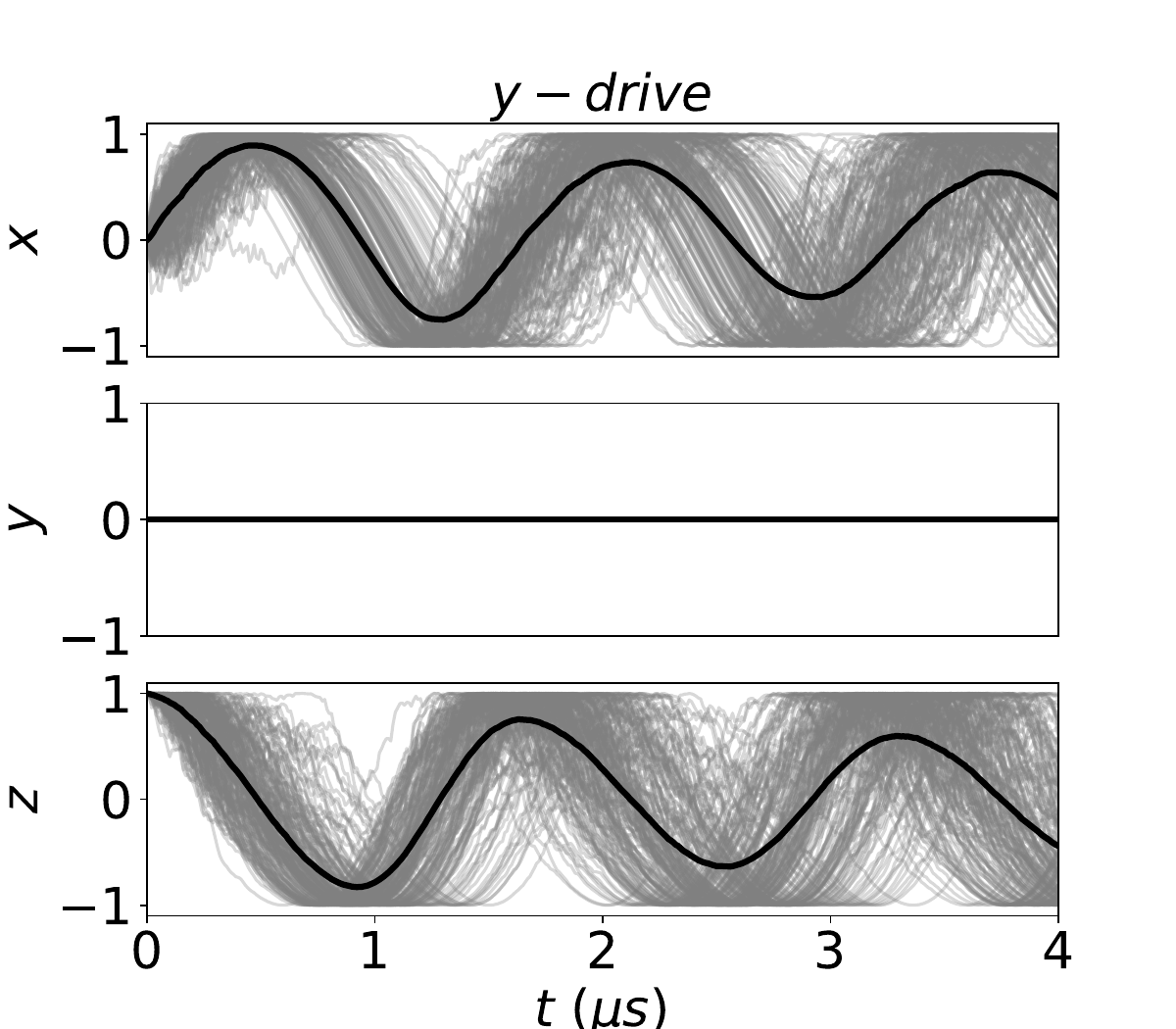}
    \includegraphics[width=0.49\linewidth]{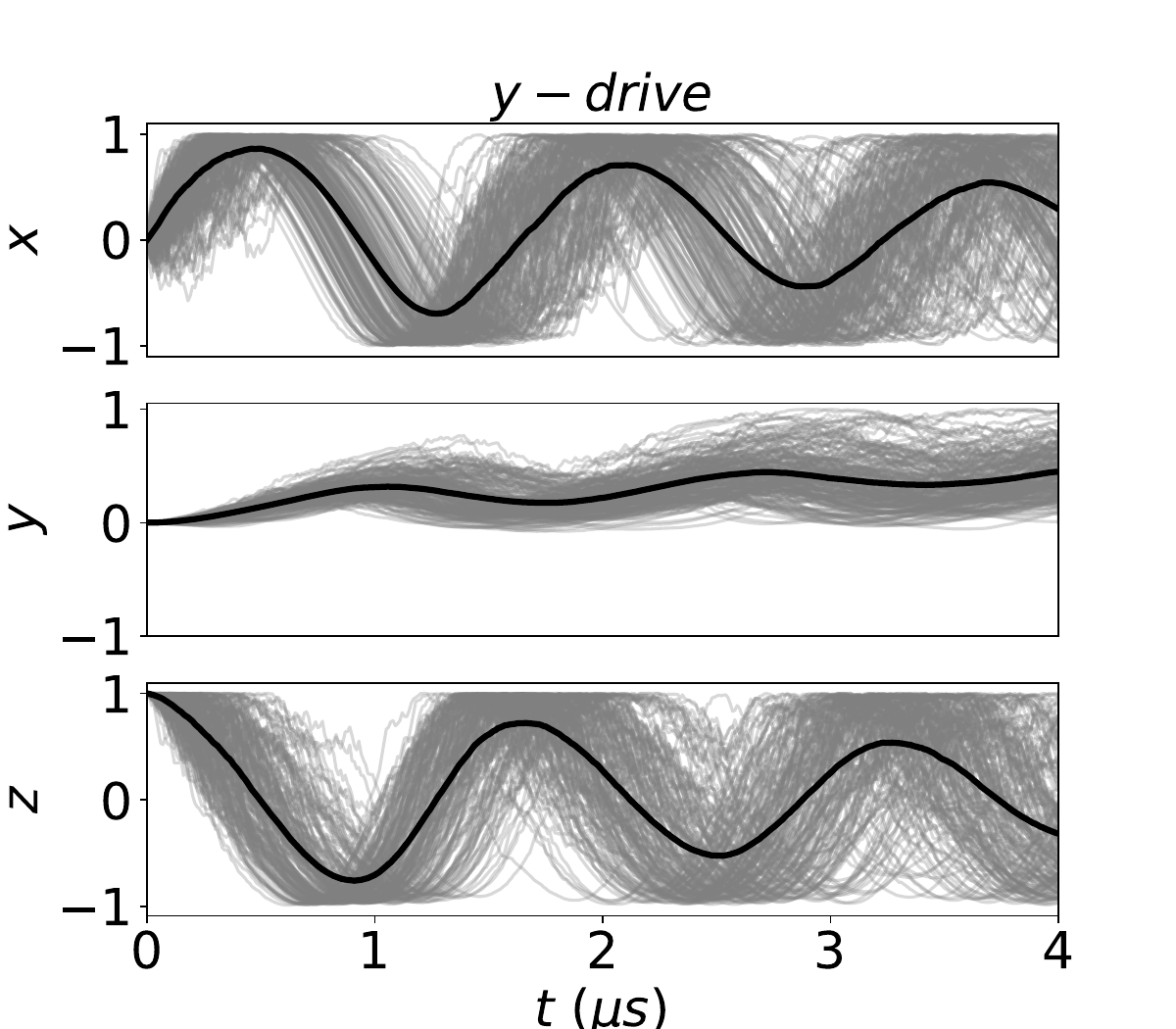}
    \caption{Bloch components trajectory evolution for $\sigma_x$-measurement wtih $\sigma_y$-drive. Black solid lines represents their respective ensemble average. Left and right column corresponds to $\Delta=0~\text{MHz}$- and $\Delta=0.5~\text{MHz}$, respectively. Other parameters are $\Gamma_g=2~\text{MHz}$, $\Gamma_e=0.2~\text{MHz}$ and $\Omega = 2~\text{MHz}$.}
    \label{fig:bloch_traj_y}
\end{figure}

 The nonlinear term (the last term in Eq.~(\ref{eq:post_SME})), which is absent in the general Lindblad dynamics, plays a significant role in the dynamics of a non-Hermitian two-level system.\cite{homodyne_roson}. We highlight the role of this non-linear term in the dynamics of the quantum battery in Sec.~\ref{sec:results}. Since the dynamics is confined to a two-level manifold, the effective density matrix can be parametrized in terms of the Bloch vector $\vec{\zeta}$ as $\rho_c = \frac{1}{2}\left(\mathbb{I} + \vec{\zeta}\cdot\boldsymbol{\sigma}\right),$ where $\boldsymbol{\sigma} = (\sigma_x,\sigma_y,\sigma_z)$ denotes the vector of Pauli matrices and $\vec{\zeta}(t)=(x(t),~y(t),~z(t))$ denotes the Bloch vector with $x(t)$, $y(t)$, and $z(t)$ representing its components.. Using Eq.~(\ref{eq:post_SME}) the ODEs for the components of the Bloch vectors in It\^{o}'s form are given as (for $\theta = 0$)

\begin{equation}
\begin{aligned}
dx ={}&\left(-\Delta y -\frac{\Gamma_g}{2}xz -\frac{\Gamma_e}{2} x \right)dt
        +\sqrt{\Gamma_e}(1+z-x^2)\,dW,
\\[2mm]
dy ={}&\left(-2\Omega z+\Delta x -\frac{\Gamma_g}{2}yz - \frac{\Gamma_e}{2}y \right)dt
        -xy\sqrt{\Gamma_e}\,dW,
\\[2mm]
dz ={}&\left(2\Omega y-\Gamma_e(1+z)+\frac{\Gamma_g}{2}(1-z^2)\right)dt
        \\ &-x(1+z)\sqrt{\Gamma_e}\,dW .
\end{aligned}
\label{eq:trajectory_eq_xdrive}
\end{equation}
for X-drive system Hamiltonian (\ref{eq: Hamiltonian}) and when
\begin{align}
H = \Delta\ket{2}\bra{2} + i\Omega(\ket{1}\bra2-\ket{2}\bra{1}),
\label{eq: y-Hamiltonian}
\end{align}

\begin{equation}
\begin{aligned}
dx ={}&\left(-\Delta y + 2\Omega z-\frac{\Gamma_g}{2}xz-\frac{\Gamma_e}{2}x\right)dt \\
        &\quad +\sqrt{\Gamma_e}(1+z-x^2)\,dW,
\\[2mm]
dy ={}&\left(\Delta x -\frac{\Gamma_g}{2}yz -\frac{\Gamma_e}{2}y\right)dt -xy\sqrt{\Gamma_e}\,dW,
\\[2mm]
dz ={}&\left(-2\Omega x -\Gamma_e(1+z)+\frac{\Gamma_g}{2}(1-z^2) \right)dt\\
        &\quad -x(1+z)\sqrt{\Gamma_e}\,dW.
\end{aligned}
\label{eq:trajectory_eq_ydrive}
\end{equation}
for the Y-drive, where we have used $r = \sqrt{\Gamma_e}\langle \sigma_x \rangle dt  + dW $ as the stochastic measurement record for each trajectory. Throughout this paper, we consider the initial state as $(x(0), y(0), z(0)) = (0,0,1)$ or $\rho(0) = \ket{2}\bra{2}$. 
In Fig.~\ref{fig:bloch_traj_x} and Fig.~\ref{fig:bloch_traj_y} we show the quantum trajectories of the Bloch-vector components along with their corresponding ensemble-averaged dynamics. For the lifted-EP case, i.e., $\Delta=0.5~\text{MHz}$, the measurement backaction is redistributed among the Bloch components. In this case the ensemble average values, $\langle x\rangle\neq0$ and $\langle y\rangle\neq0$ for the $\sigma_x$- and $\sigma_y$-drive, respectively. This leads to the charging and discharging of the quantum battery, as discussed in the following Sec.~\ref{sec:results}. 

It is important to note that, when evaluating ensemble averages, the full Kraus-operator-based Bayesian state-update approach, Eq.~(\ref{eq: state_update_equation}), should be used rather than the stochastic post-selected equation-of-motion approach, Eq.~(\ref{eq:post_SME}). In the former approach, the condition $dN=0$ is used to select the post-selected trajectories through the Bayesian state update, whereas in the latter approach, $dN=0$ is imposed directly in the equation of motion. Nevertheless, at the trajectory level, the no-jump master equation and the corresponding stochastic differential equations for the Bloch-vector components of the effective two-level system, Eqs.~(\ref{eq:trajectory_eq_xdrive}) and (\ref{eq:trajectory_eq_ydrive}), remain valid. Throughout this paper, our numerical simulations employ the Kraus-operator-based Bayesian state-update scheme to generate individual post-selected quantum trajectories.
In the following, we compare the charging dynamics and energetics of the quantum battery at the trajectory level with those obtained from the ensemble-averaged evolution, and contrast both with those of a generic two-level system. Specifically, we contrast quantities evaluated by first computing the relevant observable for each conditioned trajectory and subsequently performing the ensemble average with those obtained by evaluating the same observable using the ensemble-averaged density matrix. This comparison highlights the influence of continuous measurement and trajectory fluctuations on the charging performance of the quantum battery.

\section{Energetics of Quantum Battery: Demonic Ergotropy} 
\label{sec:energitics}
\subsection{Ergotropy}
$Ergotropy$ is a fundamental characteristic of a quantum battery and is defined as the maximum amount of work that can be extracted from a quantum state through a cyclic unitary transformation. Consider a unitary process generated by the operator $\hat{U}$. For a single realization described by the conditioned state $\rho_c$, the ergotropy is given by
\begin{align}
    \mathcal{E}_c = E[\rho_c] - \min_{\hat{U}} E[\hat{U}\rho_c\hat{U}^\dagger],
\end{align}
where $E[\rho]=\mathrm{Tr}(\rho H)$ denotes the average energy of the system. The minimization is achieved by transforming the state into its corresponding passive state, denoted by $\rho_p$. A passive state is a state from which no work can be extracted through any cyclic unitary operation. To construct the passive state, consider the spectral decompositions of the conditioned density matrix and the system Hamiltonian,
\begin{align}
    \rho_c=\sum_j p_j \ket{j}\bra{j},
\end{align}
and
\begin{align}
    H = \sum_j \epsilon_j \ket{\epsilon_j}\bra{\epsilon_j},
\end{align}
where $p_j$ and $\epsilon_j$ are the eigenvalues associated with the eigenvectors $\ket{j}$ and $\ket{\epsilon_j}$ of the density matrix and the Hamiltonian, respectively, which follow the ordering
\begin{align}
    p_1 \geq p_2 \geq \cdots,
    \qquad
    \epsilon_1 \leq \epsilon_2 \leq \cdots .
\end{align}
The passive state is then obtained by assigning the largest population to the lowest energy level, the second-largest population to the second-lowest energy level, and so on:
\begin{align}
    \rho_p=\sum_j p_j \ket{\epsilon_j}\bra{\epsilon_j}.
    \label{eq:pasive_state}
\end{align}
The ergotropy can therefore be expressed as
\begin{align}
    \mathcal{E}_c = \mathrm{Tr}(\rho_c H) - \mathrm{Tr}(\rho_p H).
    \label{eq:ergotropy_defination}
\end{align}
For continuously monitored systems, $\rho_c$ depends on the measurement record and hence varies from one trajectory to another. Consequently, the ergotropy $\mathcal{E}_c$ is itself a stochastic quantity whose value fluctuates across different quantum trajectories. For system Hamiltonian Eq. (\ref{eq: Hamiltonian}), the ergotropy can be expressed as 
\begin{align}
    \mathcal{E}_c^x(t) = \frac{\Delta}{2}z + \Omega x+ |\vec{\zeta}|\sqrt{\Omega^2 + \frac{\Delta^2}{4}},  
    \label{eq:ergo_x_bloch}
\end{align}
where $|\vec{\zeta}|= \sqrt{x^2 + y^2 + z^2}$ and the components of the Bloch vector are given by
\(
j(t)=\operatorname{Tr}[\rho_c(t)\sigma_j],
\) with $j=(x, y, z)$. A brief derivation is given in Appendix \ref{app:ergo_hamiltonian_derivation}. Similarly, for $y-$drive Hamiltonian given in Eq.~(\ref{eq: y-Hamiltonian}), we get the ergotropy expression as 
\begin{align}
    \mathcal{E}_c^y(t) = \frac{\Delta}{2}z + \Omega y+ |\vec{\zeta}|\sqrt{\Omega^2 + \frac{\Delta^2}{4}}.
    \label{eq:ergo_y_bloch}
\end{align}
Note that for continuous measurements, conditioned state is always a pure state and hence $|\vec{\zeta}|=1$. We now compare the ensemble-averaged ergotropy,
$\langle \mathcal{E}[\rho_c] \rangle$, which we refer to as the \emph{mean ergotropy}, with the ergotropy of the ensemble-averaged state,
$\mathcal{E}[\langle \rho_c \rangle]$, which we refer to as the \emph{mean-state ergotropy}. These two quantities correspond to different orders of averaging and ergotropy evaluation, since the coherence averages out to be zero in the later case and state becomes mixed. To compute $\langle \mathcal{E}[\rho_c] \rangle$, we first numerically generate the conditioned quantum trajectories, evaluate the ergotropy associated with each conditioned state $\rho_c$, and subsequently perform the ensemble average over all trajectories. In contrast, to compute $\mathcal{E}[\langle \rho_c \rangle]$, we first obtain the ensemble-averaged state by averaging over all conditioned trajectories and then evaluate the ergotropy of the resulting mixed state.
Since ergotropy is a convex functional of the quantum state, Jensen's inequality implies that \cite{Cvetkovski2012}
\begin{align}
    \langle \mathcal{E}[\rho_c] \rangle
    \geq
    \mathcal{E}[\langle \rho_c \rangle],
    \label{eq:ergotropy_inequality}
\end{align}
due to the fact that $|\vec\zeta|= \sqrt{\langle x \rangle^2+\langle y \rangle^2+\langle z \rangle^2}\leq1$. Therefore, the average amount of extractable work obtained from individual trajectories is always greater than or equal to the extractable work associated with the ensemble-averaged state due to information gained by the continuous monitoring. The mean ergotopy is also known as $Daemonic~ergotropy$ in the recent literature \cite{Francica2017, PhysRevApplied.20.044073}.   

\subsection{Instantaneous Charging Power}
Another useful characterizer of the quantum battery is the instantaneous charging power defined as the rate of change of ergotropy as a function of time, 

\begin{align}
    \mathcal{P}(t) = \frac{\mathcal{E}(t+dt)-\mathcal{E}(t)}{dt} = \frac{d\mathcal{E}(t)}{dt}.  
\end{align}

Hence, when the instantaneous charging power is positive, the battery is being charged, and for negative values, the battery is undergoing a discharging process. From Eq.~(\ref{eq:trajectory_eq_xdrive}) and $(\ref{eq:trajectory_eq_ydrive})$ along with Eq.~(\ref{eq:ergo_x_bloch}) and (\ref{eq:ergo_y_bloch}), the ensemble average of the instantaneous charging power $\langle \mathcal{P}(t)\rangle = \left\langle \frac{d\mathcal{E}(t)}{dt} \right \rangle$ can be written as 

\begin{align}
\langle \mathcal{P}_x(t) \rangle &= 
        -\frac{\Omega\Gamma_e}{2}\langle x\rangle
        -\frac{\Delta\Gamma_e}{2}\left(1+\langle z\rangle\right)
        +\frac{\Delta\Gamma_g}{4}\left(1-\langle z\rangle^2\right)
\nonumber\\
&\quad
        -\frac{\Omega\Gamma_g}{2}\langle x\rangle\langle z\rangle
        -\frac{\Delta\Gamma_g}{4}\operatorname{Var}(z)
        -\frac{\Omega\Gamma_g}{2}\operatorname{Cov}(x,z).
\label{eq:ICP_x}
\end{align}

Similarly, for the $\sigma_y$-drive,

\begin{align}
\langle \mathcal{P}_y(t) \rangle &= 
        -\frac{\Omega\Gamma_e}{2}\langle y\rangle
        -\frac{\Delta\Gamma_e}{2}\left(1+\langle z\rangle\right)
        +\frac{\Delta\Gamma_g}{4}\left(1-\langle z\rangle^2\right)
\nonumber\\
&\quad
        -\frac{\Omega\Gamma_g}{2}\langle y\rangle\langle z\rangle
        -\frac{\Delta\Gamma_g}{4}\operatorname{Var}(z)
        -\frac{\Omega\Gamma_g}{2}\operatorname{Cov}(y,z).
\label{eq:ICP_y}
\end{align}
where $\text{Var}(z)$ and $\text{Cov}(\bullet,z)$ with $\bullet=\{x,y\}$ are the variance in the $z$-coordinate of the Bloch sphere and the covariance between $\bullet$ and $z$-component, respectively. Note that these equations have been derived using the fact that $dW$ is the zero mean gaussian Wiener increment.


The instantaneous charging power discussed above for the post-selected two-level system can be compared with that of a generic two-level system (TLS) by setting $\Gamma_g=0~\text{MHz}$, such that the probability of observing the $\ket{1}\rightarrow\ket{0}$ quantum jump vanishes. In this limit, post-selection becomes redundant, and its dynamics are governed solely by the coherent drive and the continuously monitored $\ket{2}\rightarrow\ket{1}$ transition. Consequently, the ensemble average evolution of the generic two-level system (TLS) is described by the Lindblad master equation
\begin{align}
    \frac{d\rho(t)}{dt} =  -iH_{eff}\rho(t) + i\rho(t)H_{eff} + \Gamma_e\ket{1}\bra{2}\rho(t)\ket{2}\bra{1},
\end{align}
where $H_{eff}=H -i\frac{\Gamma_e}{2}\ket{2}\bra{2}$ is the effective non-Hermitian Hamiltonian for a generic two-level system. Since the $\ket{1}\rightarrow\ket{0}$ decay channel is absent, every quantum trajectory is retained in the ensemble, while the $\ket{2}\rightarrow\ket{1}$ transition is continuously monitored through homodyne detection, and hence, no post-selection is performed. In the TLS-limit, the ensemble average of instantaneous charging power is given as 
\begin{align}
    \langle \mathcal{P}_x(t)\rangle_{TLS} = -\frac{\Omega \Gamma_e}{2}\langle x \rangle - \frac{\Delta \Gamma_e}{2}(1+ \langle z \rangle),
    \label{eq:ICP_TLS_x}
\end{align}
for $\sigma_x$-drive, 
\begin{align}
     \langle \mathcal{P}_y(t)\rangle_{TLS} = -\frac{\Omega \Gamma_e}{2}\langle y \rangle - \frac{\Delta \Gamma_e}{2}(1+ \langle z \rangle),
     \label{eq:ICP_TLS_y}
\end{align}
for $\sigma_y$-drive. Hence, as discussed above, for non-zero $\Delta$, the measurement backaction is redistributed, as a result, $\langle x\rangle$ and $\langle y\rangle$ for the $\sigma_x$- and $\sigma_y$-drives, respectively, become non-zero functions of time. However, the contribution from the $\Omega$-dependent terms is comparatively smaller than that from the $\Delta$-dependent terms due to the presence of the $(1+\langle z\rangle)$ factor. Consequently, for the red-detuning case, i.e., $\Delta<0$, the instantaneous charging power becomes positive, implying charging of the battery in the TLS limit. This TLS-limit charging for $\Delta<0$ can be further understood by steady state solution of Bloch components and the steady state expression for ergotropy as discussed in Appendix \ref{app:steady_state_TLS}. In contrast, in the post-selection-based non-Hermitian case, i.e., $\Gamma_g\neq0$, the covariance and variance terms play a crucial role in determining the charging dynamics of the battery. Furthermore, for $\Delta=0~\text{MHz}$, i.e., in the exceptional-point regime, $\langle x\rangle=0$ and $\langle y\rangle=0$ for the $\sigma_x$- and $\sigma_y$-drive cases, respectively. This implies that the instantaneous charging power vanishes (vanishingly small for the $\sigma_x$-drive), resulting in neither charging nor discharging of the battery. Consequently, the ensemble-averaged ergotropy remains at its initial value.

\begin{figure}
    \centering
    \includegraphics[width=0.49\linewidth]{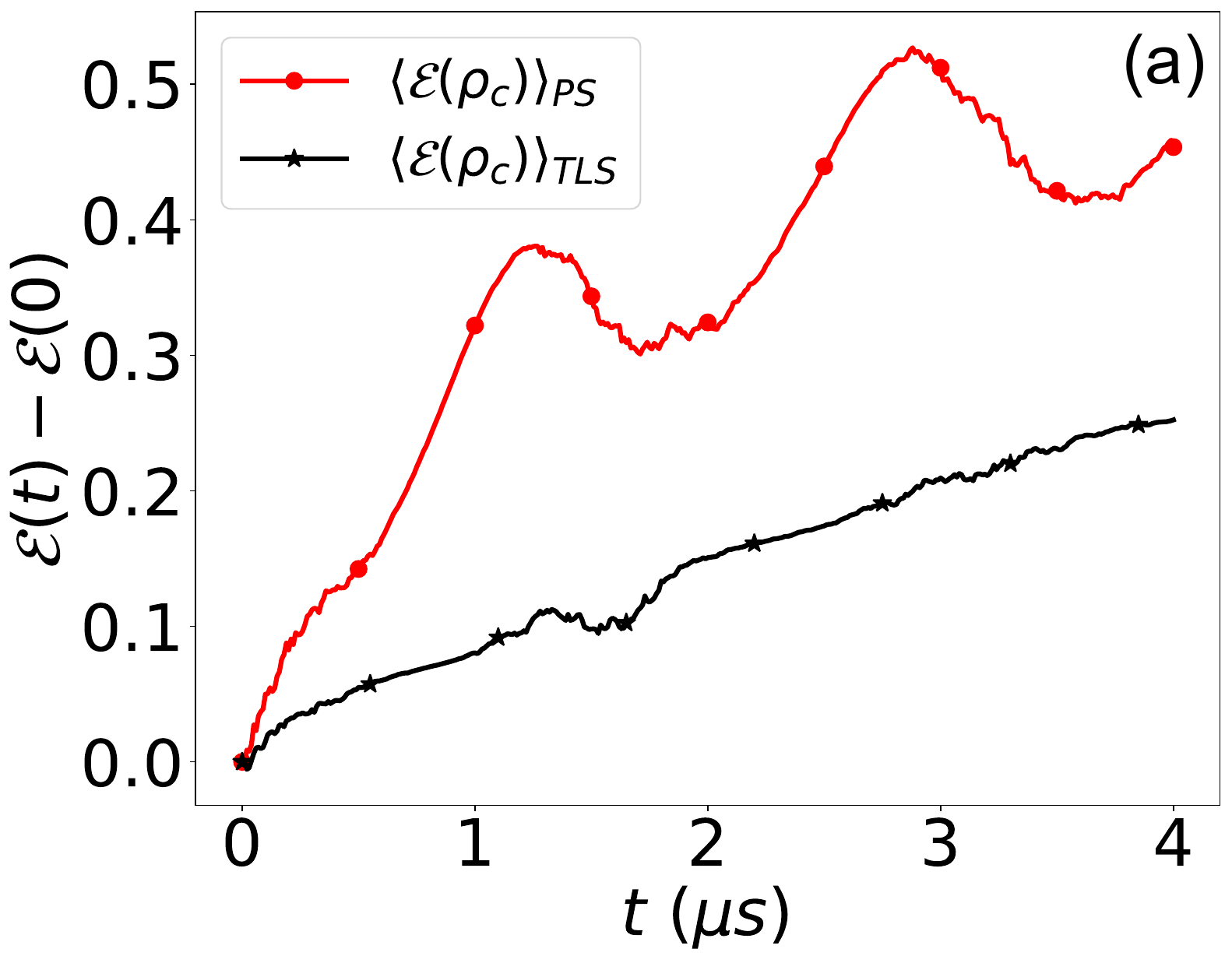}
    \includegraphics[width=0.49\linewidth]{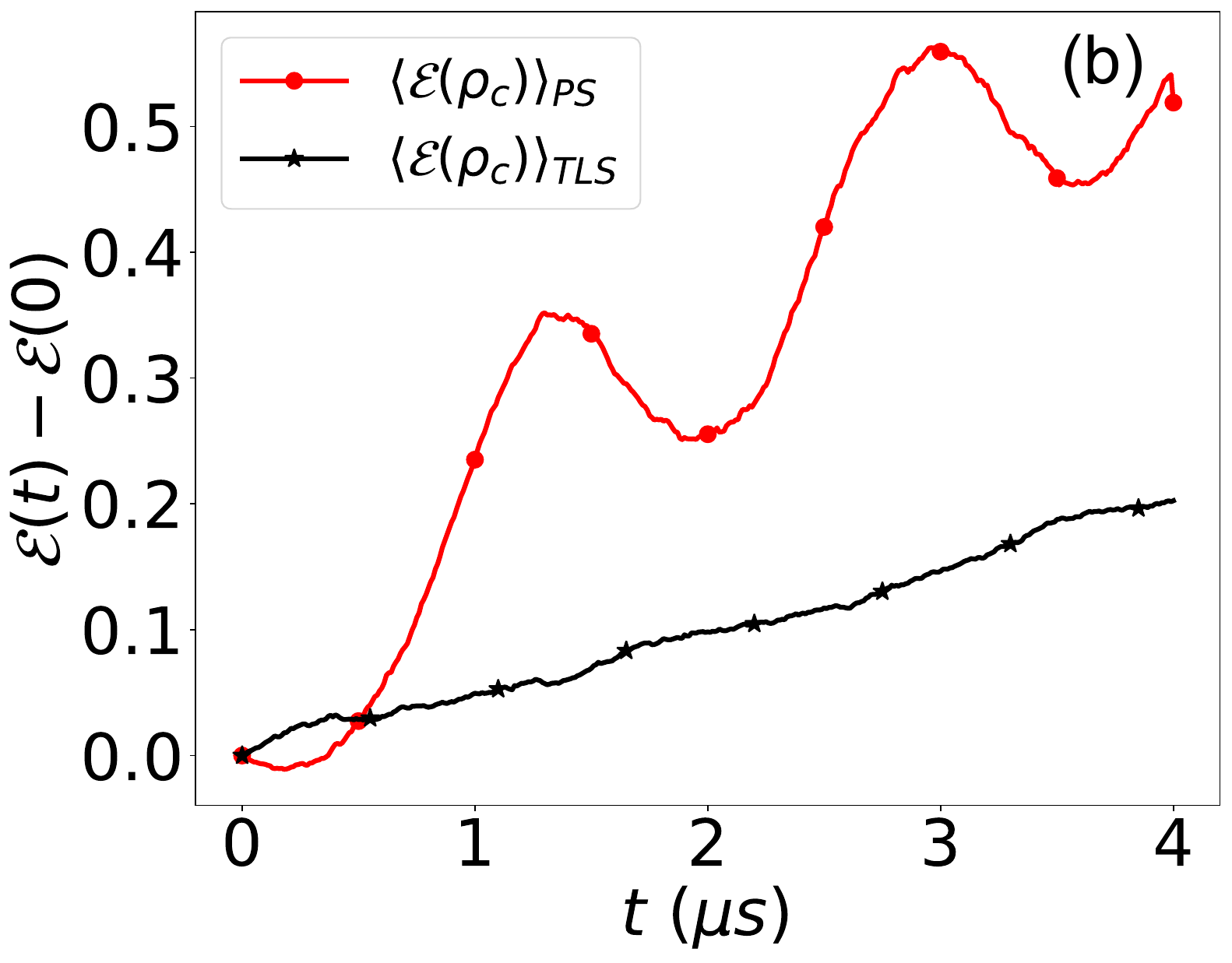}
    \caption{Ergotropy comparison between post-selected effective two level system with that of a generic two-level non-Hermitian system. (a) and (b) corresponds to $\sigma_x$- and $\sigma_y$-drive, respectively. Red and black lines corresponds to post-selected case with $\Gamma_g=2~\text{MHz},~\Delta = 0.5~\text{MHz}$ and TLS-limit case with $\Gamma_g =0,~\Delta= -0.5~\text{MHz}$, respectively. Other parameters are $\Gamma_e = 0.2~\text{MHz},~\Omega= 2~\text{MHz}$}
    \label{fig:coherence_compare}
\end{figure}

\begin{figure*}[t]
    \centering
    \includegraphics[width=0.246\linewidth]{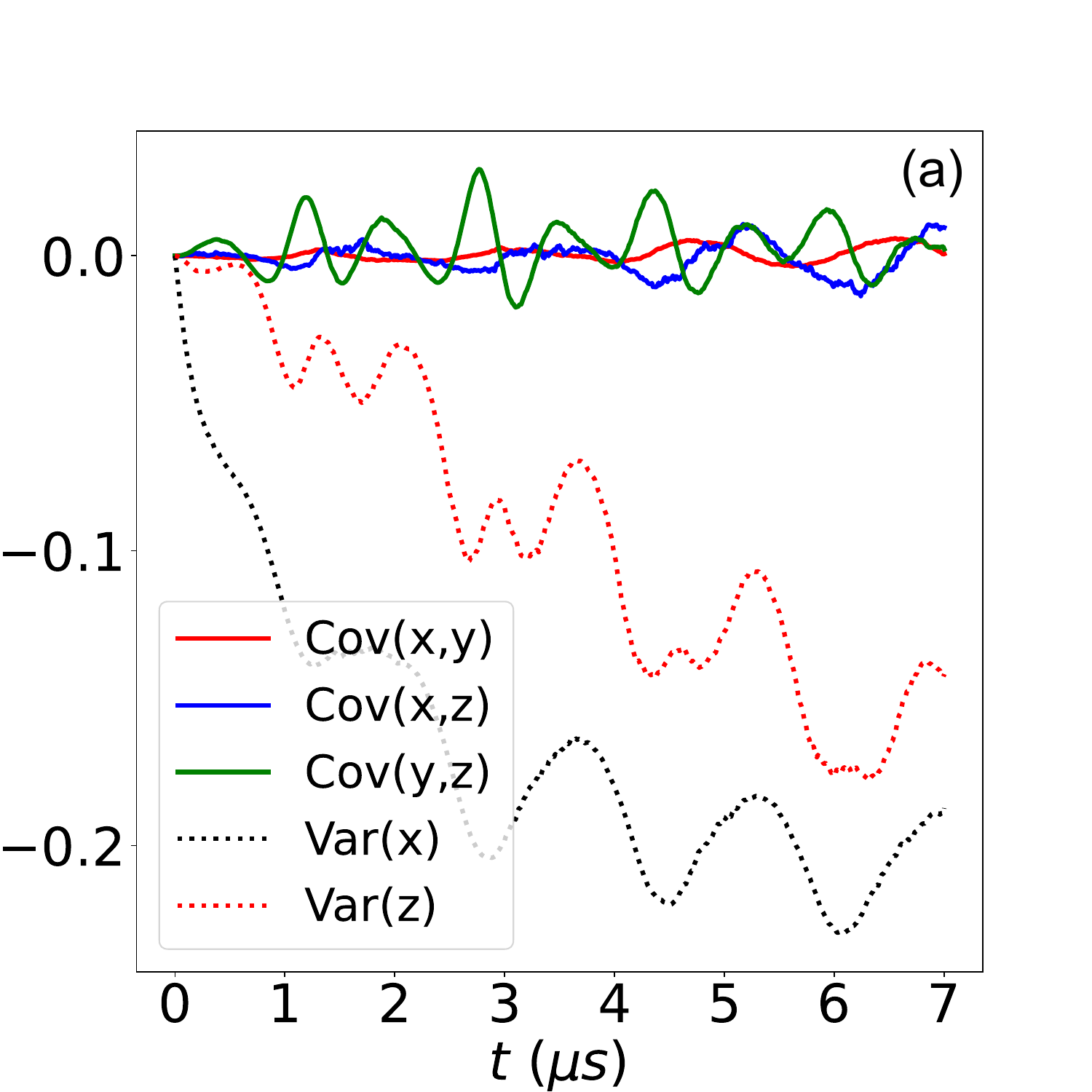}
    \includegraphics[width=0.246\linewidth]{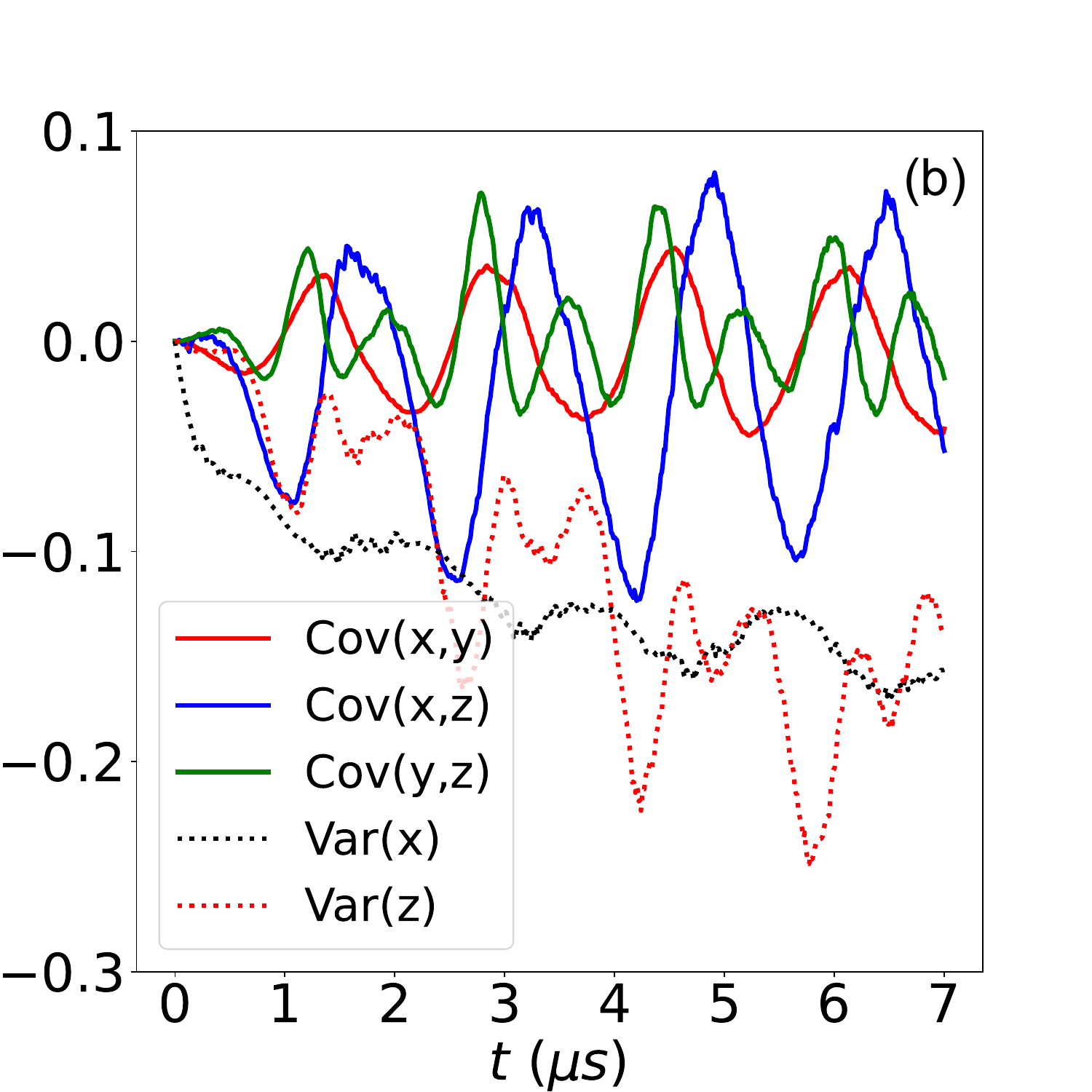}
    \includegraphics[width=0.246\linewidth]{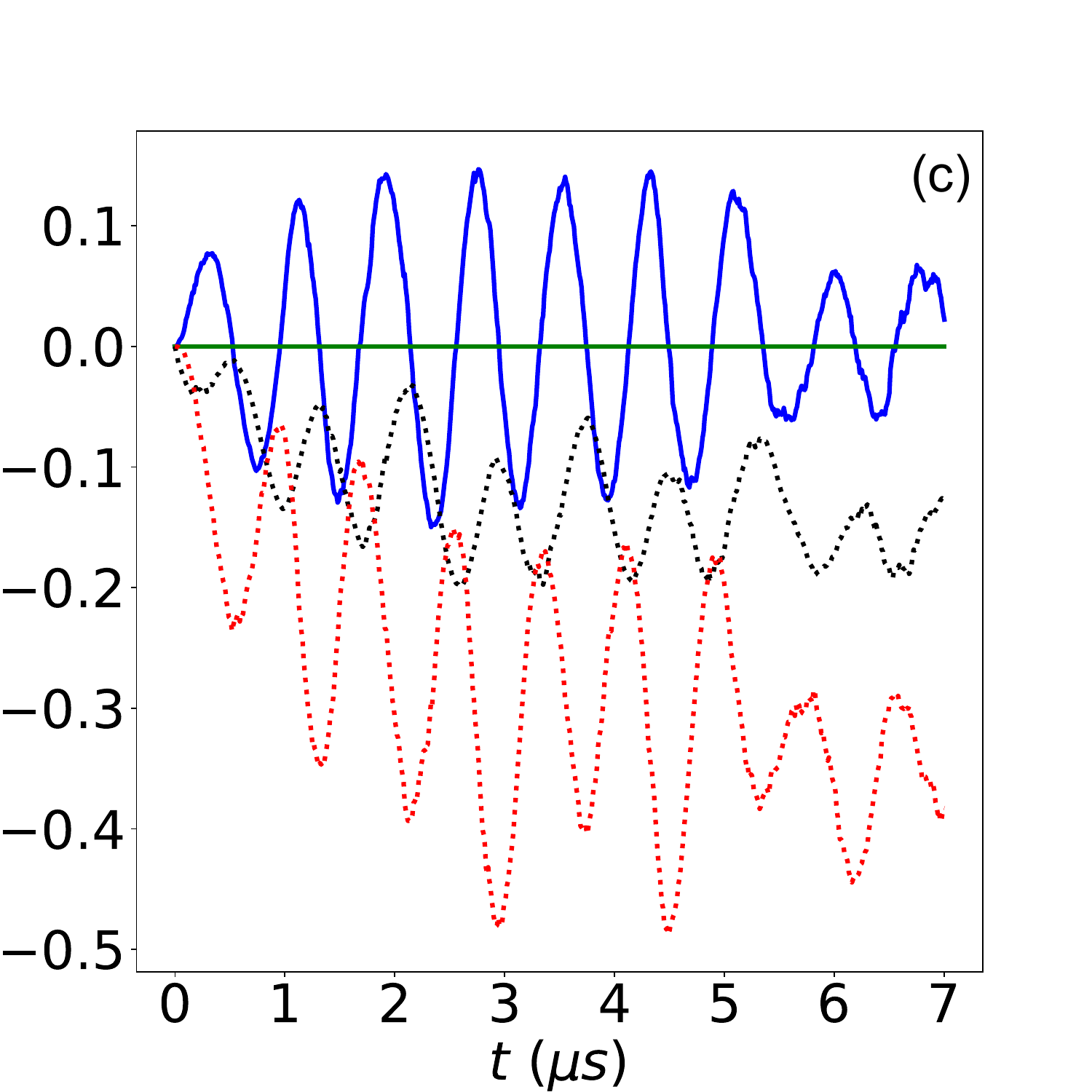}
    \includegraphics[width=0.246\linewidth]{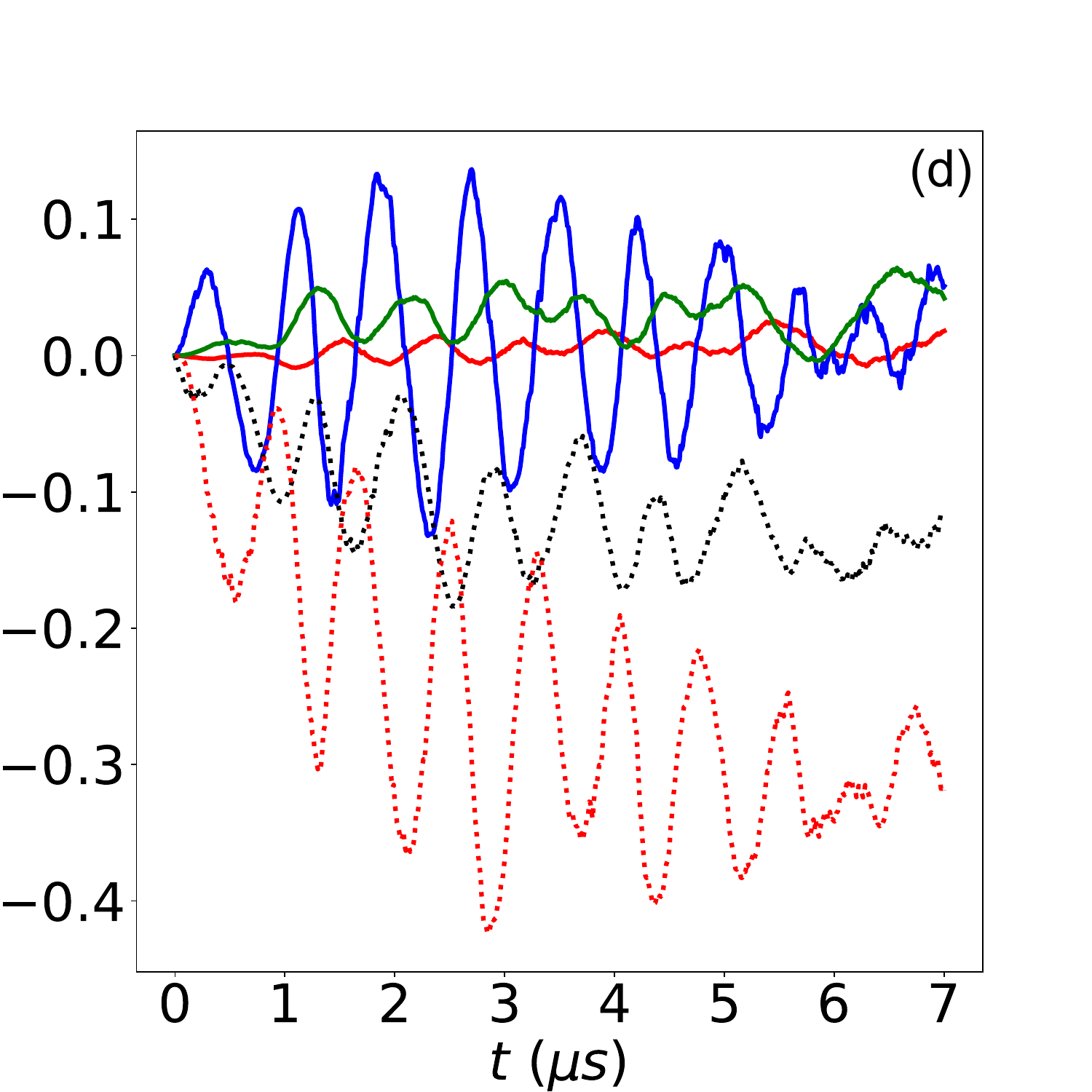}
    \caption{The covariance and variance terms associated with the Bloch components of the effective two level system. (a) and (b) corresponds to the $\sigma_x$-drive for $\Delta=0~\text{MHz}$ and $\Delta=0.5~\text{MHz}$, respectively. (c) and (d) corresponds to $\sigma_y$-drive for $\Delta=0~\text{MHz}$ and $\Delta=0.5~\text{MHz}$, respectively. Solid blue, red and green lines represents the $-\sqrt{\Gamma_e}\text{Cov}(x,y)$, $-\frac{\Gamma_g}{2}\text{Cov}(x,z)$ and $-\frac{\Gamma_g}{2}\text{Cov}(y,z)$, respectively. Whereas, dotted black and red lines represents $-\sqrt{\Gamma_e}\text{Var}(x)$ and $-\frac{\Gamma_g}{2}\text{Var}(z)$, respectively. Other parameters are $\Omega =2~\text{MHz}$, $\Gamma_g = 2~\text{MHz}$ and $\Gamma_e = 0.2~\text{MHz}$.   }
    \label{fig:Cov_Var}
\end{figure*}

\subsection{Coherence Measure} 
 Quantum coherence plays a crucial role in the charging dynamics of quantum batteries. The presence of quantum coherence has been shown to significantly influence the amount of extractable work, leading to the notions of coherent ergotropy ($\mathcal{E}_{C}$) and incoherent ergotropy ($\mathcal{E}_{IC}$) \cite{PhysRevLett.125.180603}. Coherent ergotropy represents the portion of the extractable work that is stored solely in the coherence of the quantum state. To quantify the total coherence of the battery state, we employ the $l_{1}$-norm of coherence, defined as \cite{PhysRevLett.113.140401}
\begin{align}
    C_{1}[\rho]  = \sum_{\substack{i,j\\ i\neq j}} |\rho_{ij}|,
    \label{eq:l1_coherence}
\end{align}
where $\rho_{ij}$ denotes the $(i,j)$th element of the density matrix $\rho$. Thus, the $l_{1}$-norm of coherence is simply the sum of the absolute values of all off-diagonal elements of the density matrix. For a two-level system, Eq.~(\ref{eq:l1_coherence}) reduces to
\begin{align}
    C_{1}[\rho] = \sqrt{x^{2}+y^{2}},
    \label{eq:coherence_bloch}
\end{align}
where $x$ and $y$ are the Bloch vector components. Since the $l_{1}$-norm of coherence is a convex function, it satisfies Jensen's inequality,
\begin{align}
    \sum_n p_n\, C_1[\rho_c^n] \geq C_1\!\left[\sum_n p_n \rho_c^n\right],
    \label{eq:coherence_inequality}
\end{align}
where $\rho_c^n$ is the conditioned density matrix corresponding to the $n$th quantum trajectory and $p_n$ is the probability of observing that trajectory. Thus, the ensemble average of the trajectory coherence is always greater than or equal to the coherence of the ensemble-averaged state. 

\section{Numerical Simulation Results}
\label{sec:results}

As mentioned in Sec.~\ref{sec:energitics}, to study the dynamics of the quantum battery, we first numerically employ the Kraus-operator-based Bayesian state-update equation, Eq.~(\ref{eq: state_update_equation}) or equivalently Eq.~(\ref{eq: SME}) to generate all possible realizations. Subsequently, we perform post-selection by extracting trajectories that contain no $\ket{1}\rightarrow\ket{0}$ quantum jumps. A comparison between the post-selected trajectory dynamics and the effective two-level Lindblad dynamics is presented in Appendix~\ref{app:compare_lindblad}.
  
\subsection{Post-selection Non-Hermitian Advantage} 
\label{sec:compare_two_level}

In this section, we discuss the advantage of the post-selection-based effective non-Hermitian system over a generic two-level non-Hermitian system. 
In Fig.~\ref{fig:coherence_compare}, we compare the time evolution of the ergotropy for the TLS-limit ($\Gamma_g=0~\text{MHz}$) with that of the post-selection-based non-Hermitian case ($\Gamma_g=2~\text{MHz}$). For the TLS-limit case under red detuning, $\Delta=-0.5~\text{MHz}$, and for the post-selection case under blue detuning, $\Delta=0.5~\text{MHz}$, the battery exhibits charging. However, in the post-selected case, the ergotropy dynamics are found to be highly oscillatory and attain values significantly larger than those in the TLS limit. This enhanced and oscillatory behavior arises from the covariance, variance, and other contributing terms in Eqs.~(\ref{eq:ICP_x}) and (\ref{eq:ICP_y}), which originate from the nonlinear dynamics induced by post-selection.

These results demonstrate the advantage of the post-selection-based non-Hermitian battery over a generic two-level non-Hermitian system for the same set of parameters, with a symmetric reversal of the detuning parameter. In the following sections, we investigate the ergotropy, coherence, and charging power of the post-selected two-level system in two distinct regimes: near the exceptional point (EP), $\Delta=0$ (resonant drive), and away from the EP, $\Delta\neq0$.

\subsection{$\Delta = 0$~(Exceptional Point)} 

\begin{figure}[t]
    \centering
    \includegraphics[width=0.49\linewidth]{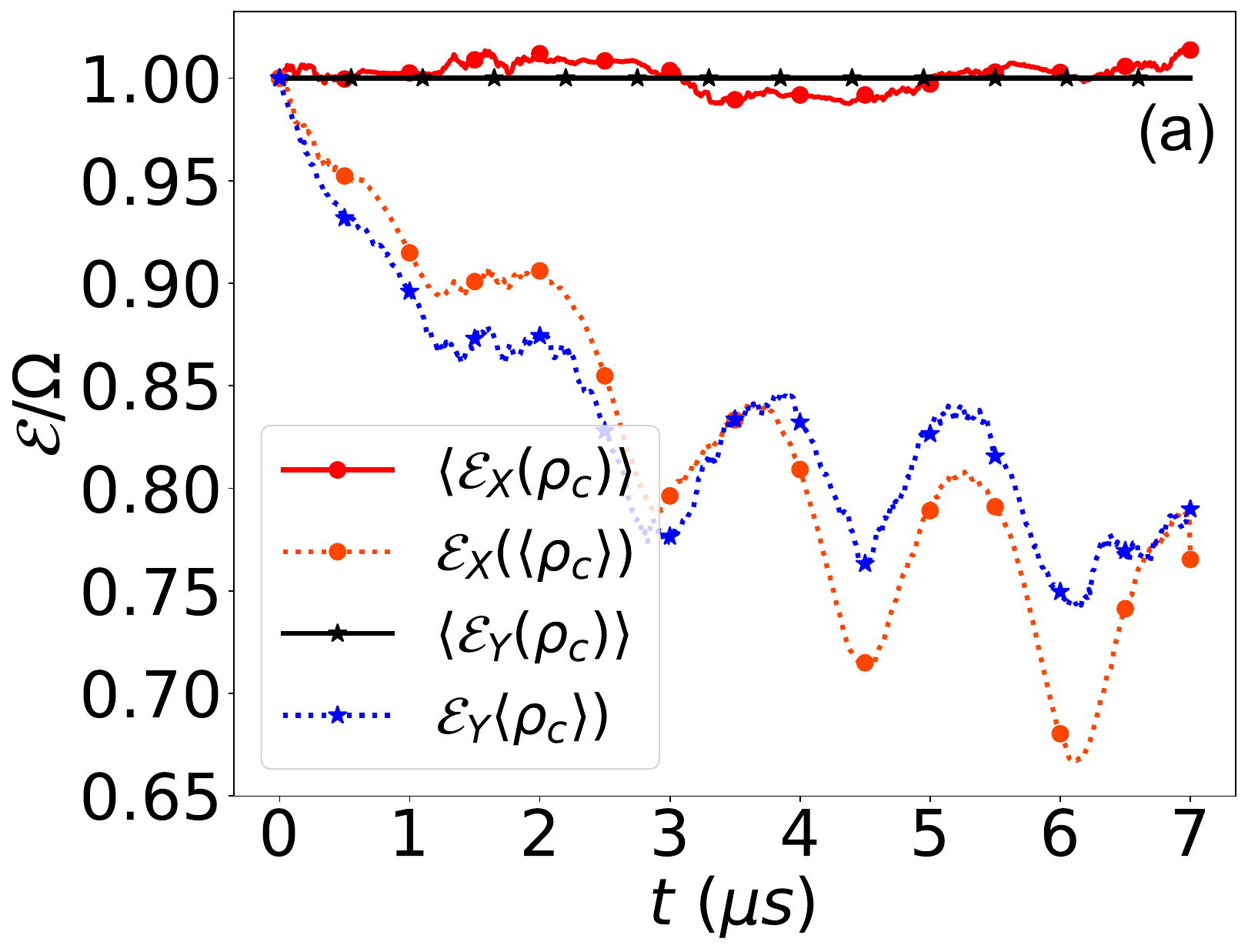}
    \includegraphics[width=0.49\linewidth]{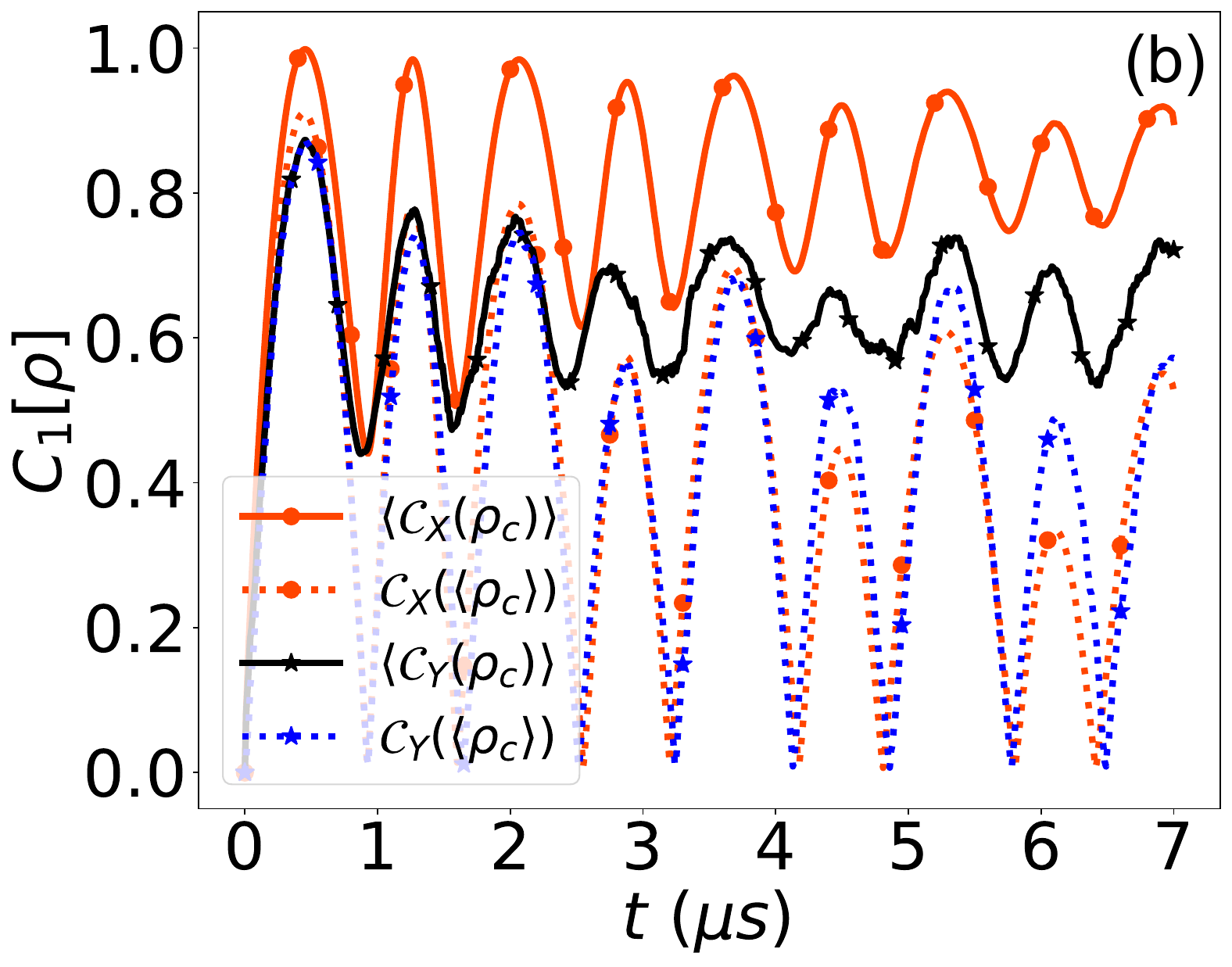}
    \includegraphics[width=0.49\linewidth]{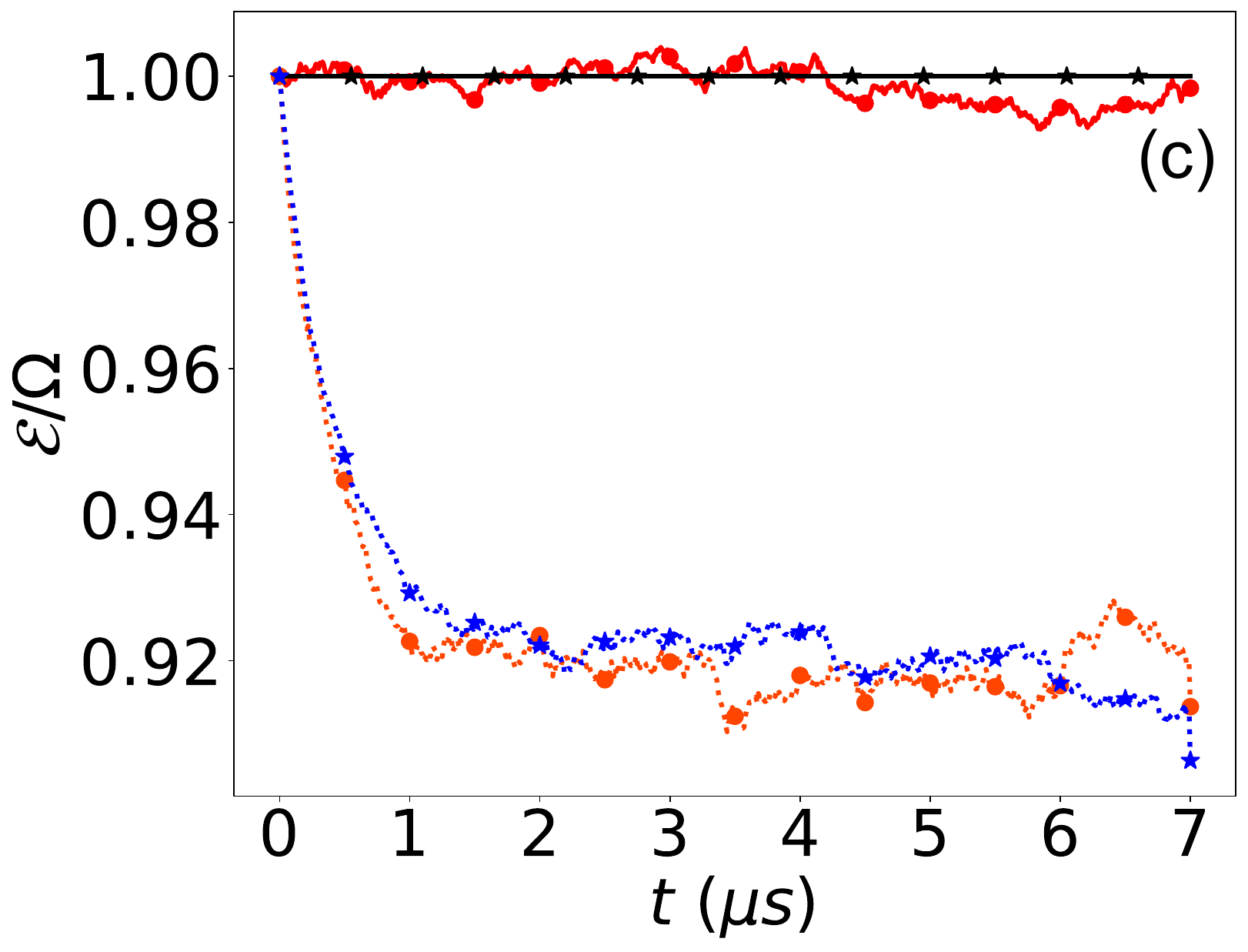}
    \includegraphics[width=0.49\linewidth]{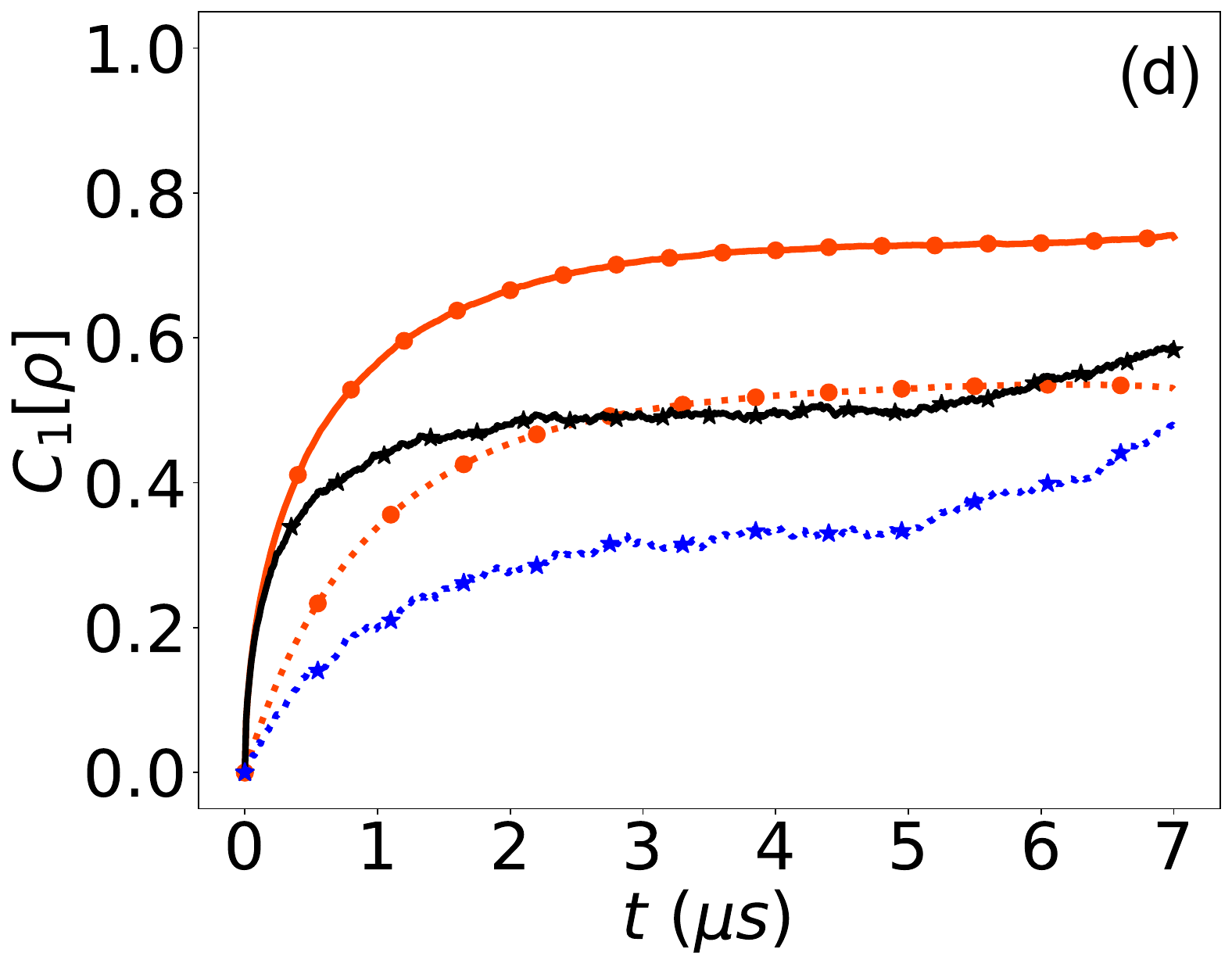}
    \caption{Ergotropy and $l_1$-norm of coherence plot. Red  and black solid lines represents mean ergotropy and mean coherence for x and y-drive, respectively for exceptional point case $\Delta =0~\text{MHz}$. Orange and blue dotted lines represents mean-state ergotropy and mean-state coherence for x and y-drive, respectively. (a) Ergotropy and (b) Coherence corresponds to $\Omega =2~\text{MHz}$. Whereas, (c) Ergotropy and (d) Coherence corresponds to $\Omega= 0.3~\text{MHz}$. Other parameters are $\Gamma_g = 2~\text{MHz}$ and $\Gamma_e =0.2~\text{MHz}$.}
    \label{fig:3}
\end{figure}

For the resonant case, \(\Delta = 0~\text{MHz}\), corresponding to \(\omega_2-\omega_1=\omega_d\), the Liouvillian spectrum exhibits an exceptional point (EP), at which the eigenvalues coalesce. We investigate the behavior of the ergotropy and coherence on the two sides of the EP, considering \(\Omega=0.3~\text{MHz}\) on the left and \(\Omega=2~\text{MHz}\) on the right, as shown in Fig.~\ref{fig:3}.
We observe that, on both sides of the EP, the mean ergotropy remains approximately constant for both the \(x\)- and \(y\)-drives, represented by the solid red and black lines, respectively. In contrast, the mean-state ergotropy exhibits qualitatively distinct behavior on the two sides of the EP. For \(\Omega=0.3~\text{MHz}\), the mean-state ergotropy decays monotonically and approximately exponentially with time. In contrast, for \(\Omega=2~\text{MHz}\), the mean-state ergotropy exhibits damped oscillations before eventually reaching a steady state. 

To understand this behavior, we note that for \(\Delta=0~\text{MHz}\), the conditioned ergotropy for the \(x\)-drive takes the simple form \(\mathcal{E}_c^x(t)/\Omega=x(t)+|\vec{\zeta}(t)|\). For an efficient continuous measurement, the conditioned state remains pure, such that \(|\vec{\zeta}|=1\). Consequently, the ergotropy approaches either its minimum or maximum value, \(\mathcal{E}_c^x/\Omega=0\) or \(\mathcal{E}_c^x/\Omega=2\) which is known as $diffusive~bifurcation$ Hence, the mean trajectory-level ergotropy remains approximately constant, \(\langle\mathcal{E}_c^x(t)\rangle/\Omega\approx1\) $(\text{since}, ~\langle x\rangle\approx 0)$, at all times.
On the other hand, in the case of $y$-drive, Bloch vector exhibits rotation in the $x$-$z$ plane.  Since the $x$-measurement projects the system into $\sigma_x$ eigenstates, the measurement basis and energy basis are now orthogonal, hence the $y$ component of Bloch vectors is always zero for all  and the ergotropy have no effect of measurement (See Fig.~\ref{fig:bloch_traj_x} and \ref{fig:bloch_traj_y}). This gives the conditioned ergotropy as $\mathcal{E}_c^y(t)/\Omega = 1$, which further implying ${\langle{\mathcal{E}}[\rho_c]\rangle}/\Omega = 1$. Hence, under continuous homodyne measurement, the mean ergotropy is preserved at its initial value despite the stochastic evolution of individual quantum trajectories. 
This can be further understood from Eqs.~(\ref{eq:ICP_x}) and (\ref{eq:ICP_y}). For $\Delta=0~\text{MHz}$, the only contributing terms are $-\frac{\Omega\Gamma_e}{2}\operatorname{Cov}(x,z)$ and $-\frac{\Omega\Gamma_e}{2}\operatorname{Cov}(y,z)$ for the $\sigma_x$- and $\sigma_y$-drives, respectively. These covariance terms have a negligibly small contribution in the resonant case (See Fig.~\ref{fig:Cov_Var}). Consequently, the instantaneous charging power is approximately zero for $\sigma_x$-drive and perfectely zero for $\sigma_y$-drive protocols, and the ensemble-averaged ergotropy remains close to its initial value. 

The mean-state ergotropy is also shown in Fig.~7. In this case, the contribution from \( |\vec{\zeta}| \) will contribute since \(x\), \(y\), and \(z\) are themselves the mean values. Although the linear part of the ergotropy will remain close to 1, the contribution from \( |\vec{\zeta}| \) will lead to damped oscillations. This contribution comes mostly from the \(y\)- and \(z\)-dynamics in the \(\sigma_x\)-drive and the \(x\)- and \(z\)-dynamics in the \(\sigma_y\)-drive. Since the mean values of \(y\) and \(z\) in the \(\sigma_x\)-drive, and \(x\) and \(z\) in the \(\sigma_y\)-drive, decay with time, the mean-state entropy will eventually decrease.

On the other hand, Fig.~\ref{fig:3}(b) and (d) show the time evolution of the ensemble-averaged $l_1$-norm of coherence, $\langle C_1[\rho_c]\rangle$, and the coherence of the ensemble-averaged state, $C_1[\langle\rho_c\rangle]$. Since the initial state has zero coherence, non-classicality begins to build due to coherent population oscillations between the $\ket{2}$ and $\ket{1}$ states, leading the system to a highly coherent state. For $\Omega=2~\text{MHz}$ (right of EP), coherence exhibits oscillatory behavior for both driving protocols. However, under the $\sigma_y$-drive, the coherence is smaller than under the $\sigma_x$-drive. 

This behavior can be understood from the dynamics of the Bloch vector components Fig.~\ref{fig:bloch_traj_x} and \ref{fig:bloch_traj_y}, together with Eq.~(\ref{eq:coherence_bloch}). In the $\sigma_y$-drive, the coherent part of the dynamics coupled to the $(x,z)$ components of the Bloch vector, and due to $x$-quadrature measurement, the measurement backaction noise enters through the $x$-component. However, in the case of $\sigma_x$-drive, the coherent Hamiltonian generates rotations in the \(y\)-\(z\) plane, while the measurement acts along the \(x\)-axis. Due to this, the measurement backaction noise gets redistributed in all components.
Therefore, for $\Delta=0~\text{MHz}$ and the $\sigma_y$-drive, the $y$ component of the Bloch vector remains zero for every realization. Consequently, only the $x$ component contributes to the $l_1$-norm of coherence, yielding $\langle C_y\rangle=\langle|x|\rangle$. In contrast, for the $\sigma_x$-drive, both the $x$ and $y$ components contribute to the coherence through $\langle C_x\rangle=\left\langle \sqrt{x^2+y^2}\right \rangle$. As a result, the coherence in the \(\sigma_y\)-drive is determined solely by the \(x\)-dynamics, whereas in the \(\sigma_x\)-drive both the \(x\)- and \(y\)-dynamics contribute to the coherence. Consequently, the coherence is larger for the \(\sigma_x\)-drive than for the \(\sigma_y\)-drive.


In the case of the mean-state coherence, \(C=\sqrt{\langle x\rangle^2+\langle y\rangle^2}\), both \(\langle x\rangle\) and \(\langle y\rangle\) decrease with time. For the \(x\)-drive, \(\langle x\rangle\approx0\), and hence there is no significant contribution from the \(x\)-component to the mean-state coherence. In contrast, for the \(y\)-drive, \(\langle y\rangle=0\), and the coherence is determined entirely by \(\langle x\rangle\). The oscillations in the nonzero components exhibit similar behavior and eventually decay with time. Consequently, the mean-state coherence decreases in both cases and eventually decays towards zero.

In both the coherence and ergotropy, the dynamics exhibit markedly different behavior for parameters on the left and right of the EP. On the left of the EP, the relevant Liouvillian eigenvalues have vanishing imaginary parts, while their real parts remain finite. In contrast, on the right of the EP, the relevant eigenvalues have both nonzero real and imaginary parts. The dynamics of the Bloch-vector components are governed by these Liouvillian eigenmodes. The real-part gap between the relevant Liouvillian eigenvalues determines the relative decay rate, while the imaginary-part gap determines the oscillation frequency. Therefore, on the left of the EP, where the imaginary parts vanish, the ergotropy and coherence exhibit no oscillatory behavior and decay monotonically. On the right of the EP, the nonzero imaginary parts give rise to oscillatory dynamics, resulting in damped oscillations in both the ergotropy and coherence a shown in Fig.~\ref{fig:3}
\subsection{$\Delta \neq0$ (Lifted-EP)}
\label{sec:lifted_EP}

In the case of non-zero $\Delta$, the Liouville eigenvalues repel and becomes complex conjugate exhibiting a lifted-EP scenario instead of exceptional point (see Fig. \ref{fig:eigenvalue_plot}). 
For $\Delta > 0$, trajectories get shifted towards the $x=1$ and $y=1$ direction for $\sigma_x$ and $\sigma_y$-drive, respectively. This $reduced~bifurcation$ can be seen in Fig.~\ref{fig:bloch_traj_x} and \ref{fig:bloch_traj_y} for $\Delta=0.5~\text{MHz}$ (right panels), leading to the charging and self-discharging of the quantum battery, as described by the mean ergotropy expression given by 
\begin{equation}
    \langle \mathcal{E}_x \rangle = \frac{\Delta}{2} \langle z\rangle + \Omega \langle x\rangle + \sqrt{\frac{\Delta^2}{4} + \Omega^2 }, 
\end{equation}
for the $x$ drive and
\begin{equation}
    \langle \mathcal{E}_y \rangle = \frac{\Delta}{2} \langle z\rangle + \Omega \langle y \rangle+ \sqrt{\frac{\Delta^2}{4} + \Omega^2 }
\end{equation}
for $y$-drive.
\begin{figure}
    \centering
    \includegraphics[width=0.49\linewidth]{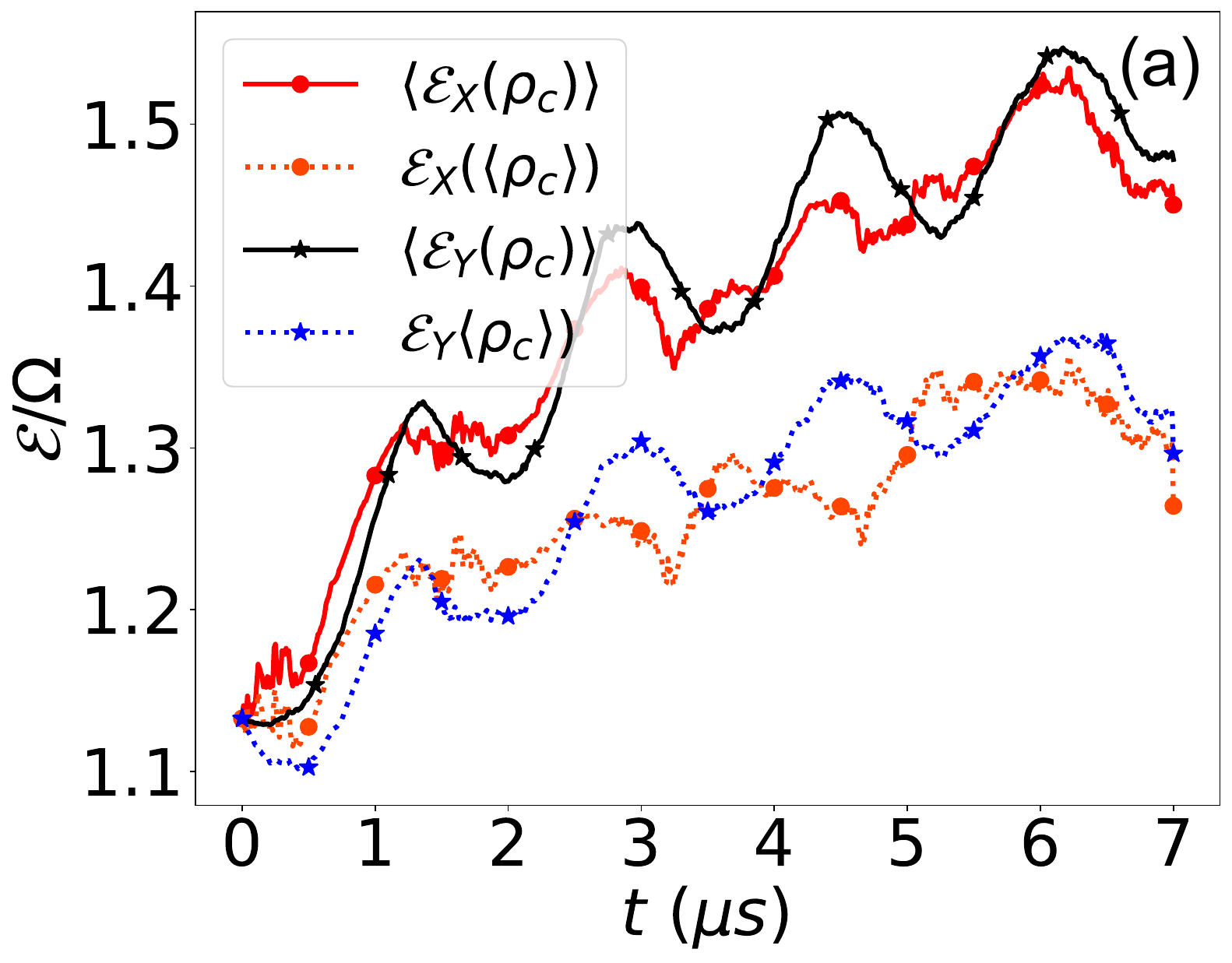}
    \includegraphics[width =0.49 \linewidth]{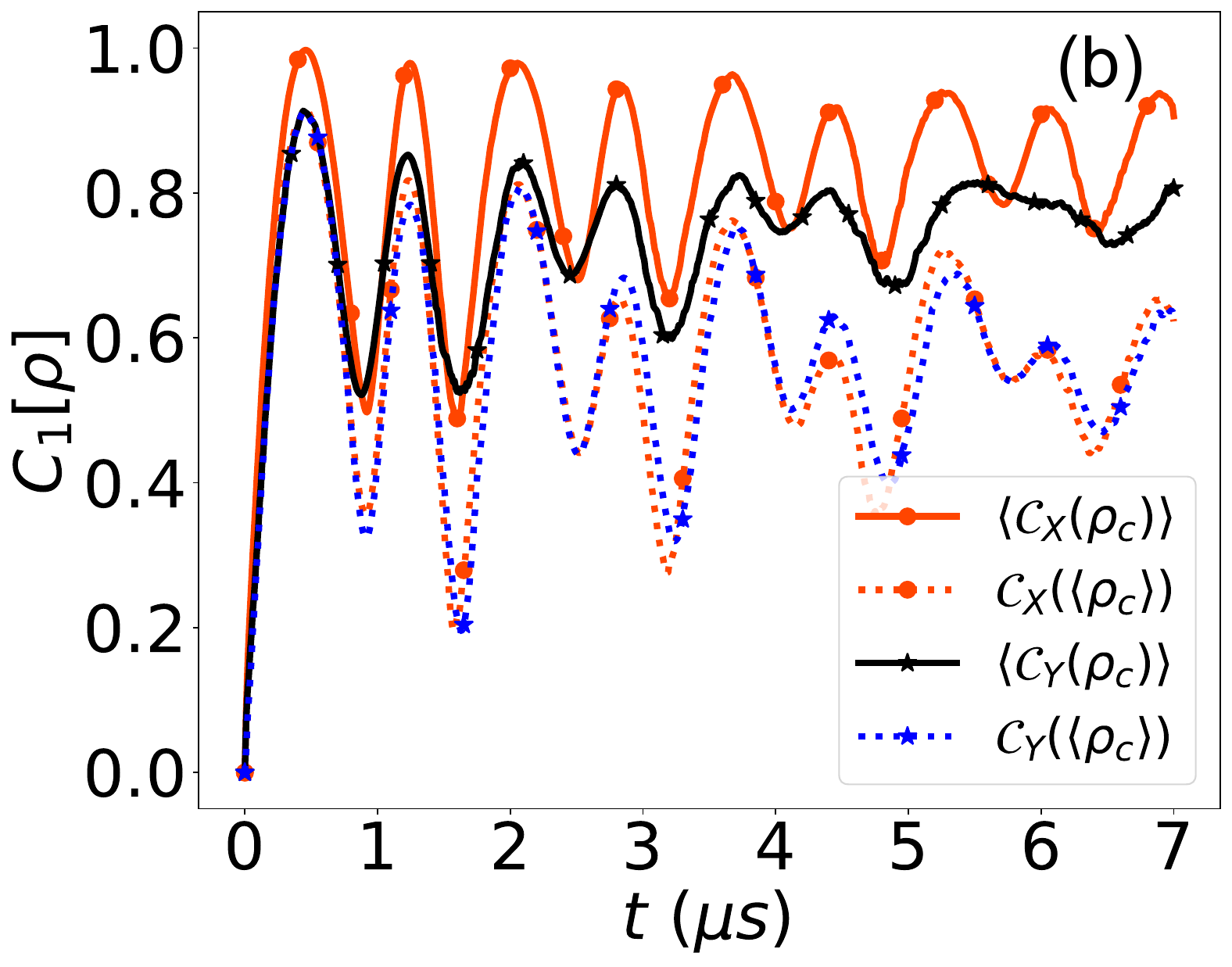} \\
    \includegraphics[width=0.49\linewidth]{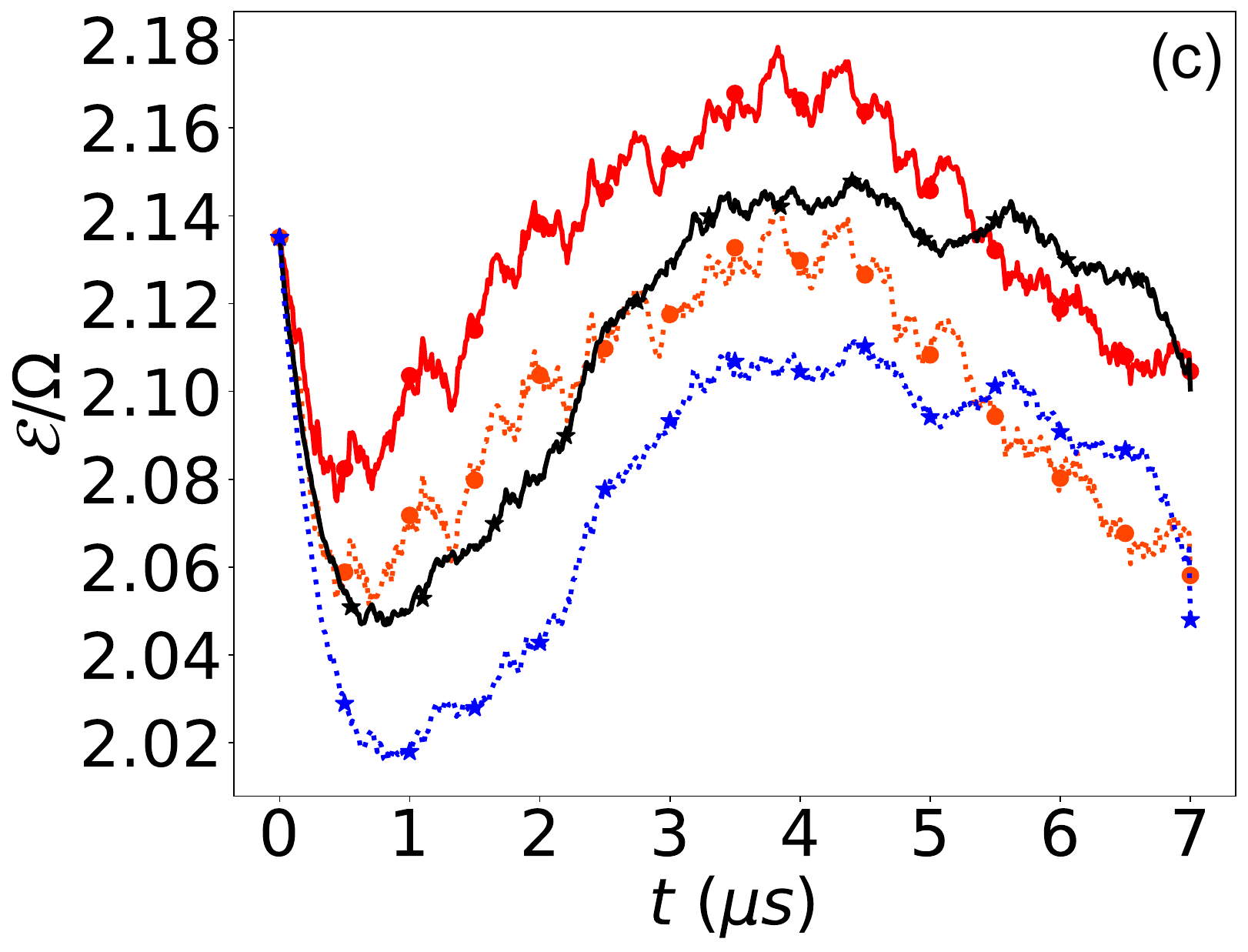}
    \includegraphics[width =0.49 \linewidth]{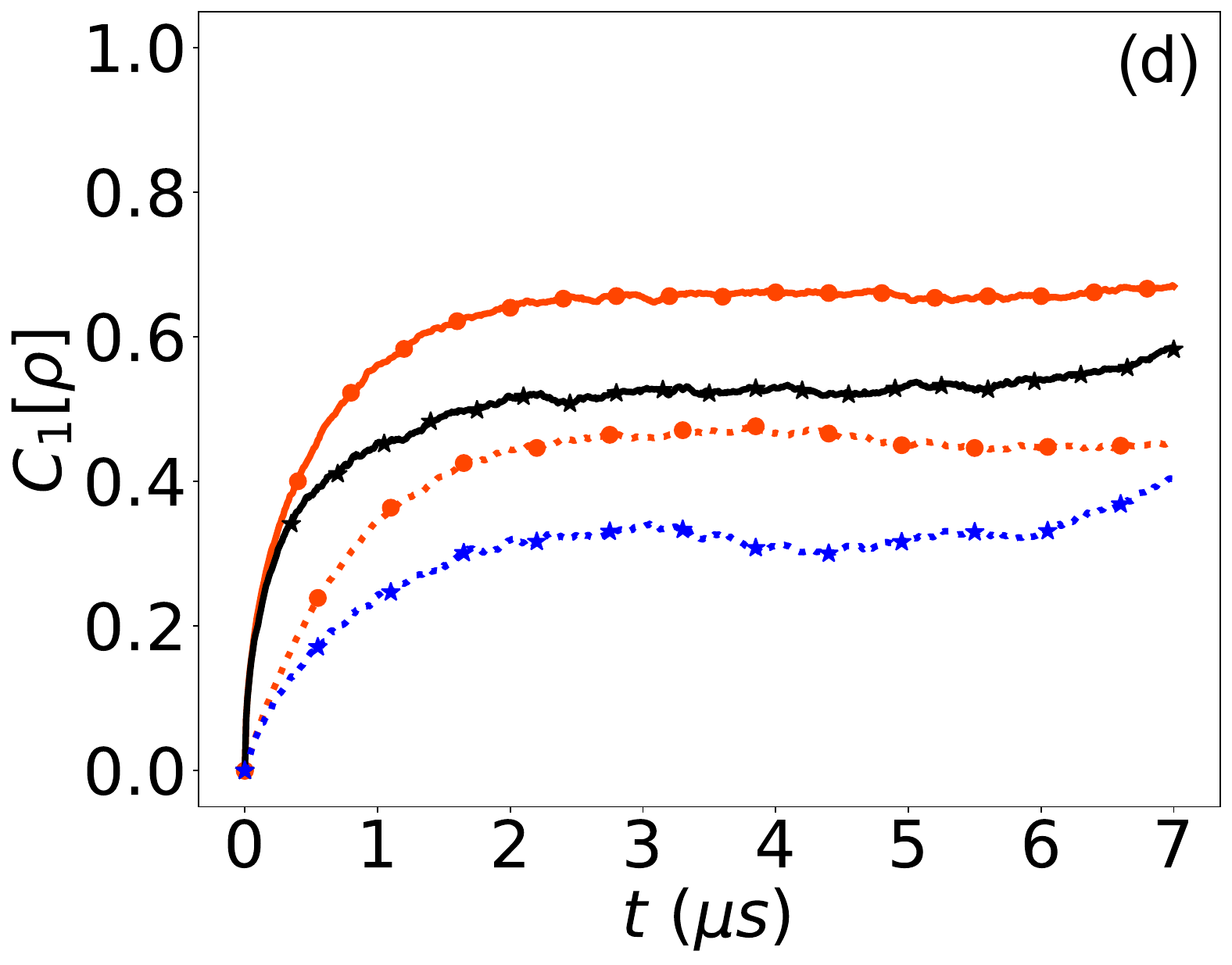}
    \caption{Ergotropy and $l_1$-norm of coherence plot. Red  and black solid lines represents mean ergotropy and mean coherence for x and y-drive, respectively for lifted-EP case $\Delta =0.5~\text{MHz}$. Orange and blue dotted lines represents mean-state ergotropy and mean-state coherence for x and y-drive, respectively. (a) Ergotropy and (b) Coherence corresponds to $\Omega =2~\text{MHz}$. Whereas, (c) Ergotropy and (d) Coherence corresponds to $\Omega= 0.3~\text{MHz}$. Other parameters are $\Gamma_g = 2~\text{MHz}$ and $\Gamma_e =0.2~\text{MHz}$.} 
    \label{fig:coherence}
\end{figure}
\begin{figure}
    \centering
    \includegraphics[width=0.49\linewidth]{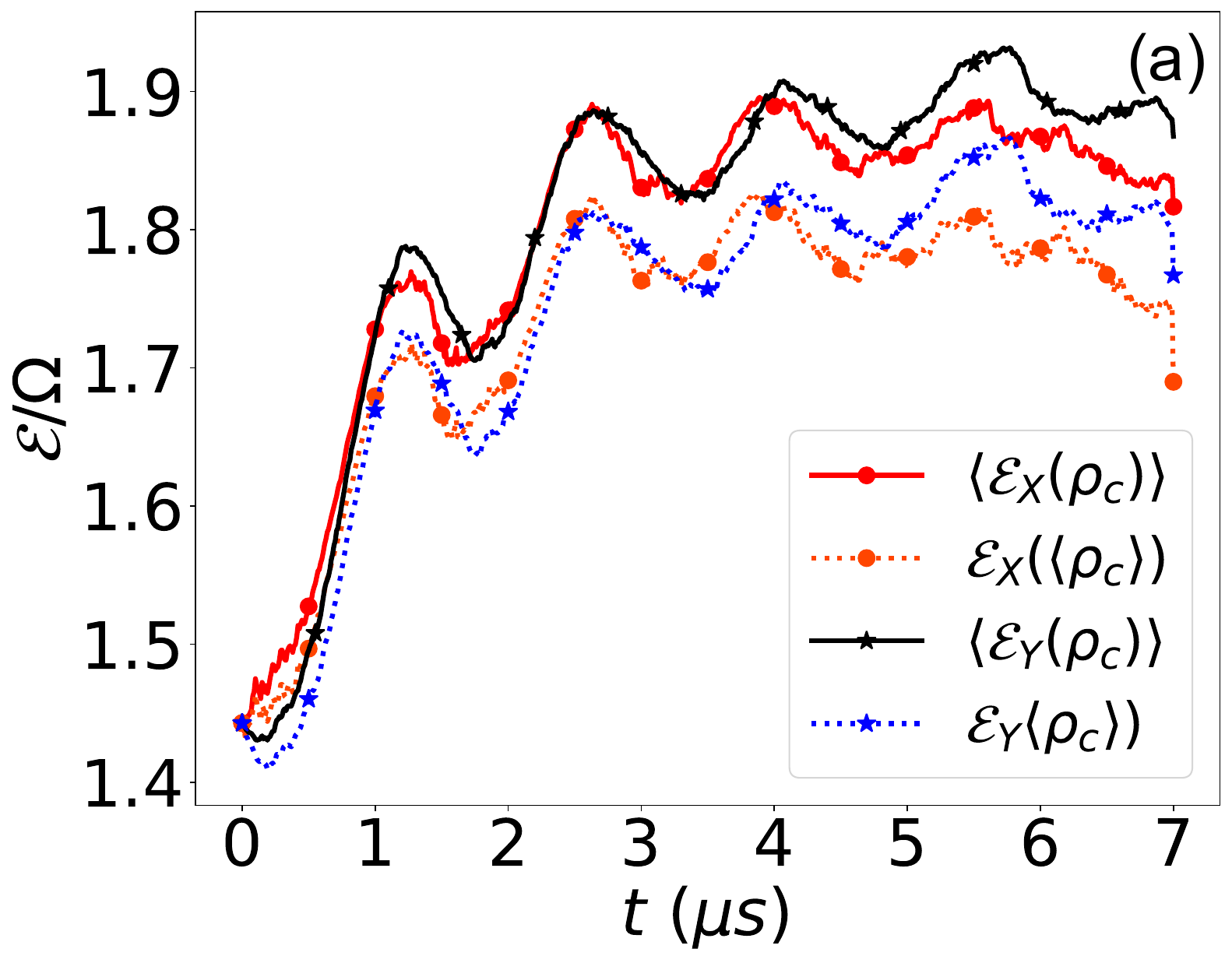}
    \includegraphics[width =0.49 \linewidth]{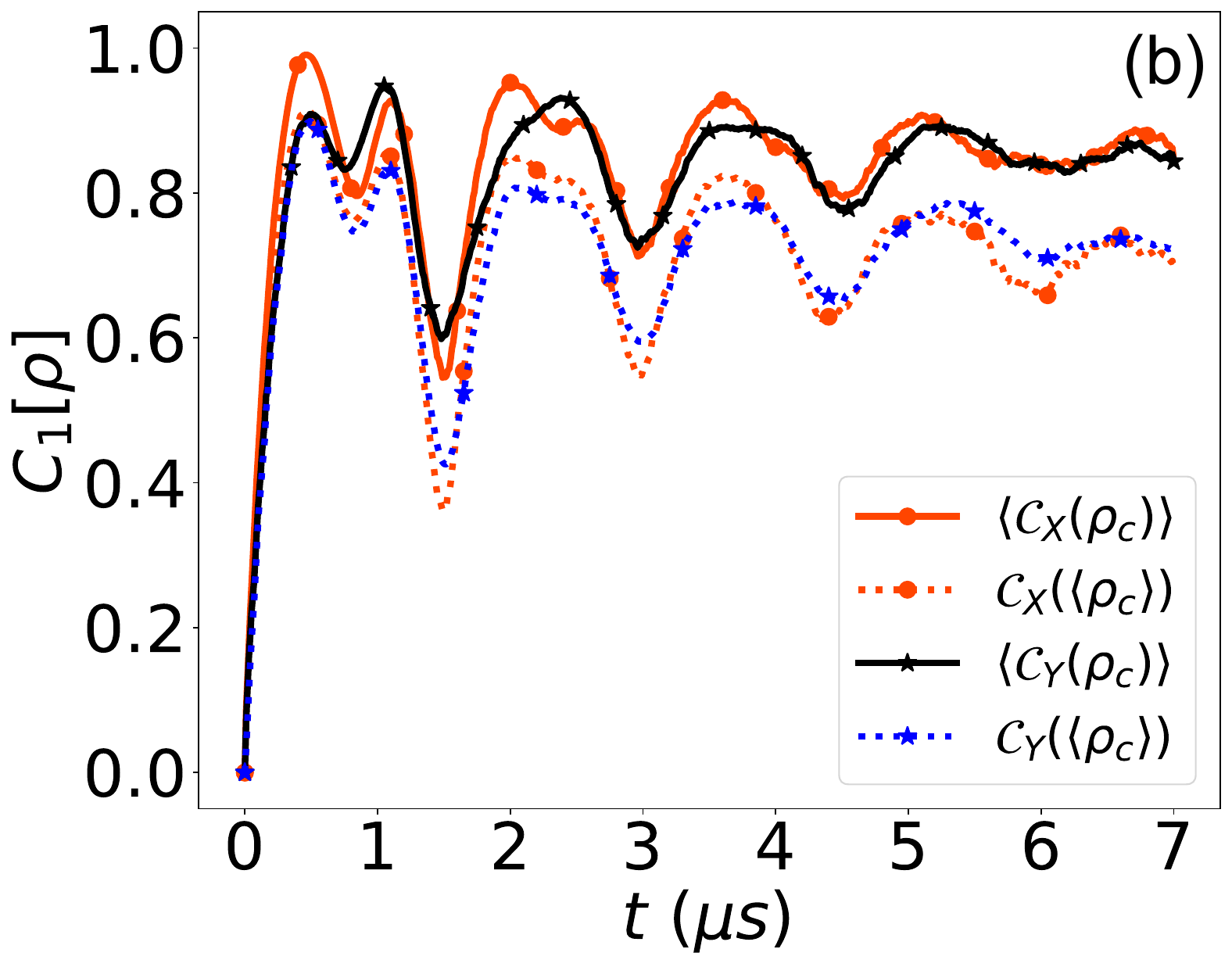} \\
    \includegraphics[width=0.49\linewidth]{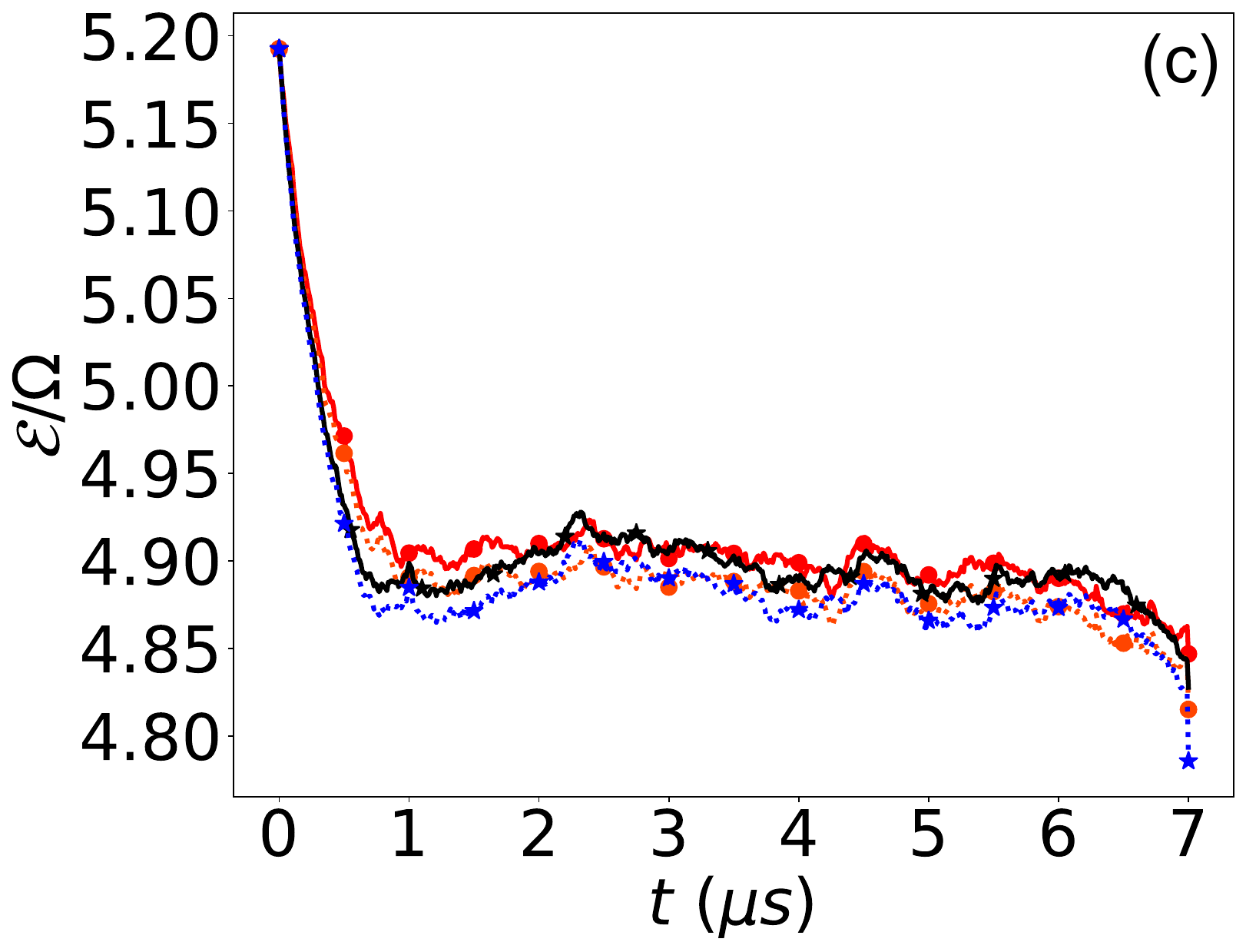}
    \includegraphics[width =0.49 \linewidth]{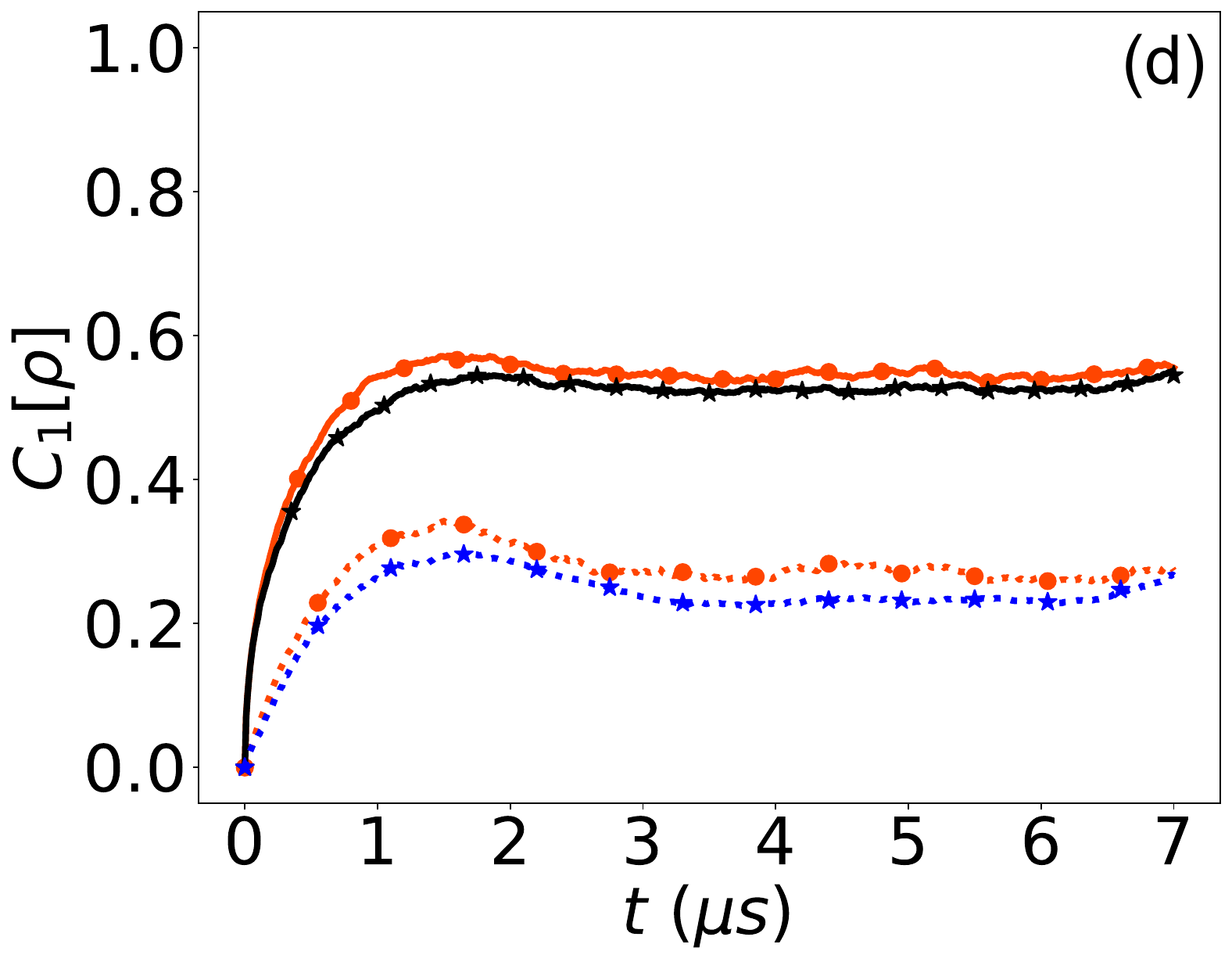} 
    
    \caption{Ergotropy and $l_1$-norm of coherence plot. Red  and black solid lines represents mean ergotropy and mean coherence for x and y-drive, respectively for lifted-EP case $\Delta =1.5~\text{MHz}$. Orange and blue dotted lines represents mean-state ergotropy and mean-state coherence for x and y-drive, respectively. (a) Ergotropy and (b) Coherence corresponds to $\Omega =2~\text{MHz}$. Whereas, (c) Ergotropy and (d) Coherence corresponds to $\Omega= 0.3~\text{MHz}$. Other parameters are $\Gamma_g = 2~\text{MHz}$ and $\Gamma_e =0.2~\text{MHz}$.}
    \label{fig:4}
\end{figure}

For non-zero detuning, the ensemble-averaged \(x\)-component for the \(\sigma_x\)-drive and the \(y\)-component for the \(\sigma_y\)-drive become non-zero. Consequently, the ergotropy receives contributions from \(\langle z\rangle\) as well as from the corresponding transverse component, \(\langle x\rangle\) or \(\langle y\rangle\), respectively.
Since these contributions are positive and increases with time, their combined effect leads to charging of the battery, resulting in an increase in the ergotropy.
In Fig.~\ref{fig:coherence}, the ergotropy and the \(l_1\)-norm of coherence are plotted as a function of time for \(\Delta=0.5~\text{MHz}\) with \(\Omega=0.3~\text{MHz}\) and \(\Omega=2~\text{MHz}\). It can be observed that both the ergotropy and coherence exhibit markedly different behavior for the two driving strengths. In particular, for \(\Omega=2~\text{MHz}\), shown in panel (a), the ergotropy exhibits oscillatory growth with time and eventually reaches a value higher than its initial value. In contrast, for \(\Omega=0.3~\text{MHz}\), the ergotropy changes only weakly with time, while the oscillatory behavior decreases. This difference can be understood from the Liouvillian spectrum, as the larger driving strength \(\Omega=2~\text{MHz}\) results in a larger separation between the imaginary parts of the relevant Liouvillian eigenvalues, leading to a higher oscillation frequency and more pronounced oscillatory dynamics. Moreover, the dynamics of \(\langle x\rangle\), \(\langle y\rangle\), and \(\langle z\rangle\) will have contributions from the variance of \(z\) and the covariances of \(xz\) and \(yz\), as can be seen from the It\^o differential equations, thereby contributing to the ergotropy. Since the covariances and variances for the \(\sigma_y\)- and \(\sigma_x\)-drives are different, as shown in Fig.~\ref{fig:Cov_Var}, the deviation in ergotropy is also expected to be different, as discussed in Appendix~\ref{app:compare_lindblad}. In Fig.~\ref{fig:4}, it can be seen that, for large values of \(\Delta\), the deviations in the \(\sigma_y\)- and \(\sigma_x\)-drives become comparatively small, and hence the deviations in coherence and ergotropy also decrease.

\begin{figure} 
    \centering
    \includegraphics[width=1.1\linewidth]{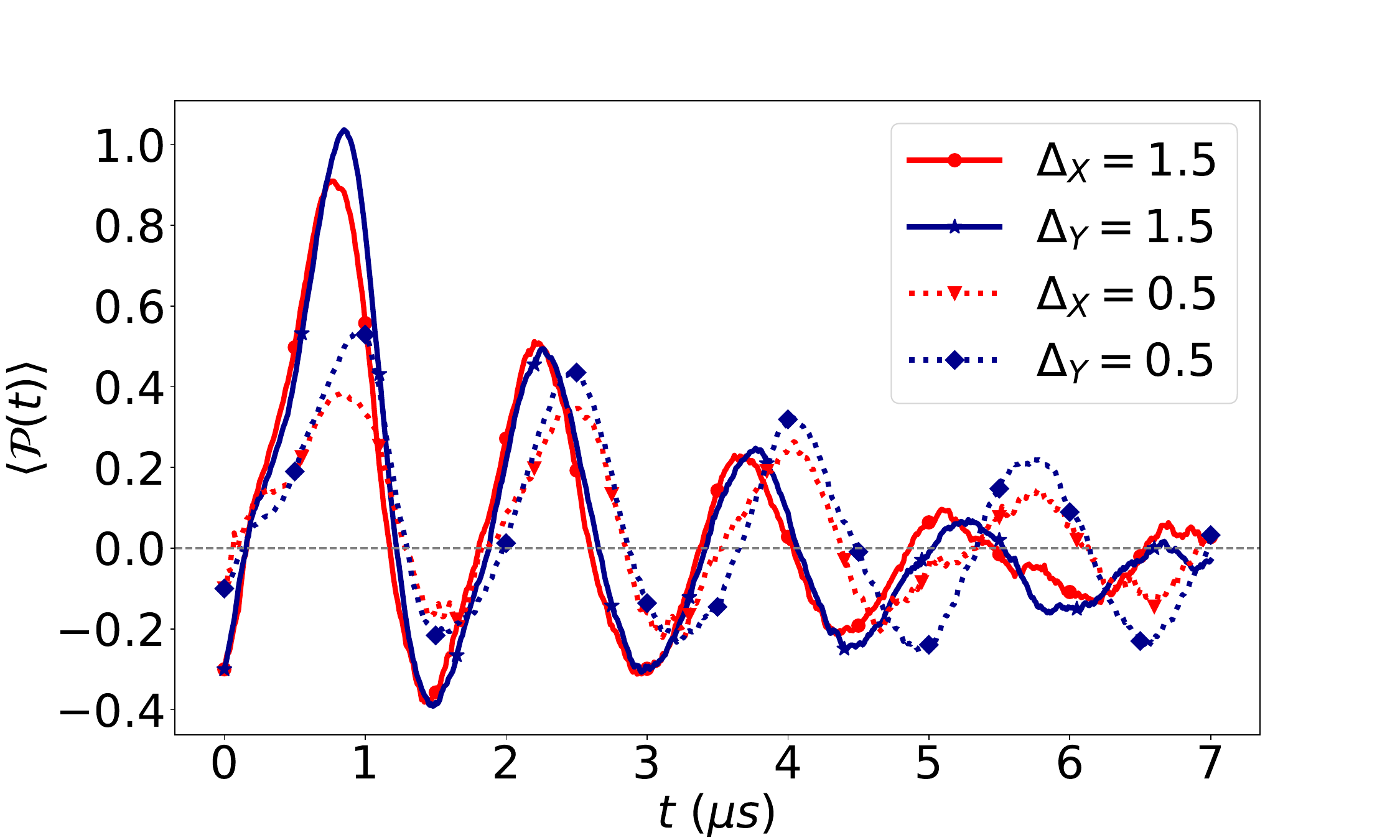}
    \caption{Instantaneous charging power as a function of time. Red and blue lines corresponds to $\sigma_x$ and $\sigma_y$-drive, respectively. Solid lines represents $\Delta =1.5~\text{MHz}$ case, whereas, $\Delta=0.5~\text{MHz}$ case represented by dotted lines. Other parameters are $\Omega=2~\text{MHz}$, $\Gamma_g = 2~\text{MHz}$ and $\Gamma_e = 0.2~\text{MHz}$}
    \label{fig:avg_power_battery}
\end{figure}

The charging dynamics discussed above can be further understood through the instantaneous charging power. Fig.~\ref{fig:avg_power_battery} shows the instantaneous charging power for different values of $\Delta$, calculated using Eqs.~(\ref{eq:ICP_x}) and (\ref{eq:ICP_y}) for $\Omega =2~\text{MHz}$. It can be observed that, for $\Delta=1.5~\text{MHz}$, the instantaneous charging power is larger than that for $\Delta=0.5~\text{MHz}$ during the initial stage of evolution, indicating a higher rate of accumulation of extractable work. However, for $\Delta=1.5~\text{MHz}$, the charging power decays more rapidly than for $\Delta=0.5~\text{MHz}$. 
The corresponding behavior of the ergotropy is compared in Fig.~\ref{fig:3} and Fig.~\ref{fig:4}, where a larger value of \(\Delta\) leads to faster saturation of the ergotropy than a smaller value of \(\Delta\).
The faster decay for larger values of \(\Delta\) can be explained in terms of the Liouvillian spectrum. As \(\Delta\) increases, the gap between the real parts of the relevant Liouvillian eigenvalues becomes larger relative to the gap between their imaginary parts. This leads to a faster decay of the charging power.

\section{Discussion and Conclusion }
\label{sec:conclusion}

In conclusion, we have investigated the dynamics of a non-Hermitian quantum battery realized through continuous quantum measurements and post-selection of trajectories with no $\ket{1}\rightarrow\ket{0}$ quantum jumps. We showed that for finite detuning, the exceptional-point degeneracy in the Liouvillian spectrum is lifted, giving rise to an lifted-EP regime. We systematically analyzed the interplay between the detuning, driving strength, and driving axis, and demonstrated how these parameters govern the charging dynamics and the extractable work of the highly tunable non-Hermitian quantum battery. We further investigated the role of detuning $\Delta$ in the coherence dynamics of the post-selected non-Hermitian system. Our results confirm that, for larger $\Delta$, both coherence and ergotropy saturate much faster than for smaller $\Delta$. This behavior is a direct consequence of the increase in the Liouvillian spectral gap. Consequently, the coherence and ergotropy reach their respective steady-state values at earlier times as the detuning is increased.
 
We also compared the dynamics of the post-selection-based non-Hermitian quantum battery with those of a continuously monitored generic two-level non-Hermitian quantum battery. We found that ergotropy exhibits markedly distinct behavior in the two cases. For the TLS-limit case, the ergotropy increases monotonically, whereas in the post-selection case, the battery undergoes oscillatory charging. This oscillatory behavior is a direct consequence of the non-linearity in the equations of motion arising from post-selection and normalization under continuous monitoring of the system. These results demonstrate a clear advantage of the effective two-level quantum battery realized from a three level system via post-selection over its generic two-level counterpart in terms of energy storage and work extraction. We have also compared the trajectory-based dynamics with the corresponding Liouvillian dynamics. Due to the nonlinear normalization inherent in the conditioned trajectory dynamics, the two descriptions generally deviate from each other. We find that this deviation is reduced as the Liouvillian spectral gap increases, which can be achieved by increasing either the driving amplitude or the detuning. These findings highlight the role of measurement backaction and non-Hermitian dynamics together governs the dynamics of quantum battery and provide a pathway for
exploring controlled charging and work extraction near exceptional points. 



\section*{Acknowledgment}
A.V. gratefully acknowledges a research fellowship from MoE, This work is supported by MoE, Government of India (Grant No. MoE-STARS/STARS-2/2023-0161). Government of India. 

\appendix

\section{Liouville Super-Operator} 
\label{appendix A}
Here we give details of the spectral properties of the Liouville super operator. The general Lindblad master equation for the reduced system density matrix is given as \cite{10.1093/acprof:oso/9780199213900.001.0001} 
\begin{align}
    \frac{d\rho(t)}{dt} = -i[H, \rho(t)] + \left(L\rho(t)L^\dagger - \frac{1}{2}\{L^\dagger L, \rho(t)\}\right)
\end{align}
where $L$ is the Lindblad operator. This equation can be further written in the super operator form by vectorizing the density operator as 
\begin{align}
    \partial_t\vec \rho_t = {\overline{\overline{\mathcal{L}}}}\vec \rho_t.
    \label{eq:eigenvalues_equation}
\end{align}
Here, the Liouville super operator is given as 
\begin{align}
    {\overline{\overline{\mathcal{L}}}} = -i(H \otimes I-I\otimes H^T) + L \otimes L^* - \frac{L^\dagger L\otimes I+I\otimes L^T L^*}{2}
\end{align}
where superscript $T$ and $*$ represents transpose and complex conjugate. Now the eigenvalues of  the Liouville super-operator are written in the form of left-handed and right-handed eigenvalues 
\begin{align}
   {\overline{\overline{\mathcal{L}}}}\vec{\phi}_i = \lambda_i\vec{\phi}_i \\ 
   {\overline{\overline{\mathcal{L}}}}^\dagger\vec{\psi}_i = \lambda_i^*\vec{\psi}_i
\end{align}

where $\vec{\phi}$ and $\vec{\psi}$ are the left-handed and right-handed eigenvectors, respectively, and $\lambda_i$ are the eigenvalues of the Liouville super-operator. The super-operator form of Liouville operator in EQ.~(\ref{eq:Lindblad_ME}) is given as 

\begin{align}
{\overline{\overline{\mathcal{L}}}}
=
\begin{pmatrix}
-\Gamma_e & -i\Omega  & i\Omega & 0 
\\[3pt]
-i\Omega & -\dfrac{\Gamma_e+\Gamma_g}{2}-i\Delta & 0 & i\Omega
\\[3pt] 
i\Omega & 0 & -\dfrac{\Gamma_e+\Gamma_g}{2}+i\Delta & -i\Omega 
\\[3pt]
\Gamma_e & i\Omega & -i\Omega & 0
\end{pmatrix}.
\label{eq: Liouville_super}
\end{align}
The eigenvalues of this matrix are ploted in Fig.~(\ref{fig:eigenvalue_plot}). 


\section{Hamiltonian and Ergotropy Expression derivation}
\label{app:ergo_hamiltonian_derivation}
In the appendix, we will provide a short derivation of Hamiltonian Eq.~(\ref{eq: Hamiltonian}) and then using this Hamiltonian we will derive the ergotropy relation Eq.~(\ref{eq:ergo_x_bloch}). For a driven three level system (Fig.~\ref{fig: model}), the Hamiltonian is given as 
\begin{align}
    H = H_0 + H_d
\end{align}
where $H_0$ and $H_d$ are the system Hamiltonian and $H_d$ is the Hamiltonian of the external $\sigma_x$ drive applied between the states $\ket{2}$ and $\ket{1}$, and are given as 
\begin{align}
    H_0 = \omega_2 \ket{2}\bra{2} + \omega_1 \ket{1}\bra{1},
    \\
    H_d = \Omega(e^{-i\omega_d t}\ket{2}\bra{1} + e^{i\omega_d t}\ket{1}\bra{2}) 
\end{align}

where $\Omega$ is the strength of the external derive. Transforming the total Hamiltonian into the rotating frame of the external drive with the unitary operator $U= e^{i\omega_d t}$. The transformed Hamiltonian is given as 
\begin{align}
    H' = UHU^\dagger + i \frac{dU}{dt}U^\dagger
\end{align}
we get 
\begin{align}
    H' = \left((\omega_2- \omega_1)- \omega_d\right) \ket{2}\bra{2} + \Omega (\ket{2}\bra{1}+ \ket{1}\bra{2}). 
    \label{eq:app_hamiltonian}
\end{align}
Here, the detuning is defined as $\Delta = (\omega_2- \omega_1)- \omega_d$. The eigenvalues and eigenvecotrs of this Hamiltonian is given as 
 \begin{align}
     E_{\pm} = \frac{\Delta\pm\sqrt{\Delta^2 + 4\Omega^2}}{2}, 
     \\
     \ket{E_{\pm}} = \mathcal{N}_\pm
     \begin{bmatrix}
        \frac{\Delta \pm \sqrt{\Delta^2 + 4\Omega^2}}{2\Omega} \\ 1
     \end{bmatrix}
 \end{align}

where $\mathcal{N}_\pm$ are the normalization factors. On the other hand, the eigenvalues of density matrix represented in terms of Bloch components are given as  

\begin{align}
    \rho_{\pm} = \frac{1\pm |\vec \zeta|}{2}
\end{align}
where $|\zeta | = \sqrt{x^2 + y^2 + z^2}$ is defined as the length of the Bloch vector. Hence, using Eq.~(\ref{eq:pasive_state}), the passive state is given as 

\begin{align}
    \rho_p = \frac{(1+|\vec\zeta|)}{2} \ket{E_-}\bra{E_-} + \frac{(1-|\vec\zeta|)}{2} \ket{E_+}\bra{E_+} 
\end{align}
After simple algebra and using Eq.~(\ref{eq:ergotropy_defination}), ergotropy can be expressed as 
\begin{align}
    \mathcal{E}_c^x(t) = \frac{\Delta}{2}z + \Omega x+ |\vec{\zeta}|\sqrt{\Omega^2 + \frac{\Delta^2}{4}}.   
\end{align}
Similar calculations can be performed with $\sigma_y$-drive to get Eq.~(\ref{eq:ergo_y_bloch}). 


\section{Steady State Solution and Ergotropy in TLS-Limit}
\label{app:steady_state_TLS}
The Ito form of stochastic ODEs of Bloch components in TLS-limit ($\Gamma_g=0$) corresponding to $\sigma_x$-drive are given as 
\begin{multline}
    dx = \left(-\Delta y - \frac{\Gamma_e}{2}x\right)dt + \sqrt{\Gamma_e}(1+z-x^2)dW, 
    \\
    dy = \left( -2\Omega z + \Delta x - \frac{\Gamma_e}{2}y \right)dt - xy\sqrt{\Gamma_e} dW, 
    \\
    dz = \left(2\Omega y- \Gamma_e(1+z)\right)dt - (x(1+z))\sqrt{\Gamma_e}dW.
\end{multline}

The mean steady state solutions of these ODEs are given as (using $\langle dW\rangle =0$), 

\begin{align}
    \langle x\rangle_{ss} = -\frac{8\Omega\Delta}{8\Omega^2 + 4 \Delta^2+ \Gamma_e^2}, 
    \\
    \langle y\rangle_{ss} = \frac{4\Omega\Gamma_e}{8\Omega^2 + 4 \Delta^2+ \Gamma_e^2},
    \\
    \langle z\rangle_{ss} = -\frac{{4\Delta^2}+ {\Gamma_e^2}}{8\Omega^2 + 4 \Delta^2+ \Gamma_e^2} 
\end{align}

Hence, for $\Delta <0$ leads to $\langle x\rangle_{ss}>0$ and $\langle z \rangle_{ss}<0$. Similarly for the $\sigma_y$-drive case, It\^{o} ODEs in TLS-limit are given as 

\begin{multline}
    dx = \left(-\Delta y + 2\Omega z- \frac{\Gamma_e}{2}x\right)dt + \sqrt{\Gamma_e}(1+z-x^2)dW, 
    \\
    dy = \left(+ \Delta x - \frac{\Gamma_e}{2}y \right)dt - xy\sqrt{\Gamma_e} dW, 
    \\
    dz = \left(-2\Omega x- \Gamma_e(1+z)\right)dt - (x(1+z))\sqrt{\Gamma_e}dW.
\end{multline}

The steady state solutions are given as 

\begin{align}
    \langle x \rangle_{ss} = - \frac{4\Gamma_e \Omega}{8\Omega^2 + 4\Delta^2 + \Gamma_e^2}, 
    \\
    \langle y\rangle_{ss} = -\frac{8\Omega\Delta}{8\Omega^2 + 4\Delta^2 + \Gamma_e^2}
    \\
    \langle z \rangle_{ss} = -\frac{4\Delta^2 + \Gamma_e^2}{8\Omega^2 + 4\Delta^2 + \Gamma_e^2}
\end{align}
Hence, for $\Delta<0$, $\langle y\rangle_{ss}>0$ and $\langle z \rangle_{ss}<0$. The steady state ergotropy with respect to the initial state is then given as 

\begin{align}
\delta\langle\mathcal{E}\rangle
&= \langle\mathcal{E}\rangle_{\mathrm{ss}}
-\langle\mathcal{E}\rangle(0) \nonumber\\
&= -\frac{\Delta\left(16\Omega^2+4\Delta^2+\Gamma_e^2\right)}
{8\Omega^2+4\Delta^2+\Gamma_e^2}.
\label{eq:delta_ergotropy}
\end{align}
for both driving protocols. Hence, the observation that $\delta\langle\mathcal{E}\rangle>0$ for $\Delta<0$ indicates that the battery undergoes charging for negative detuning, while it undergoes discharging for positive detuning in the generic two-level non-Hermitian case in the long time-limit.

\begin{figure}[t]
    \centering
     \includegraphics[width=0.48\linewidth]{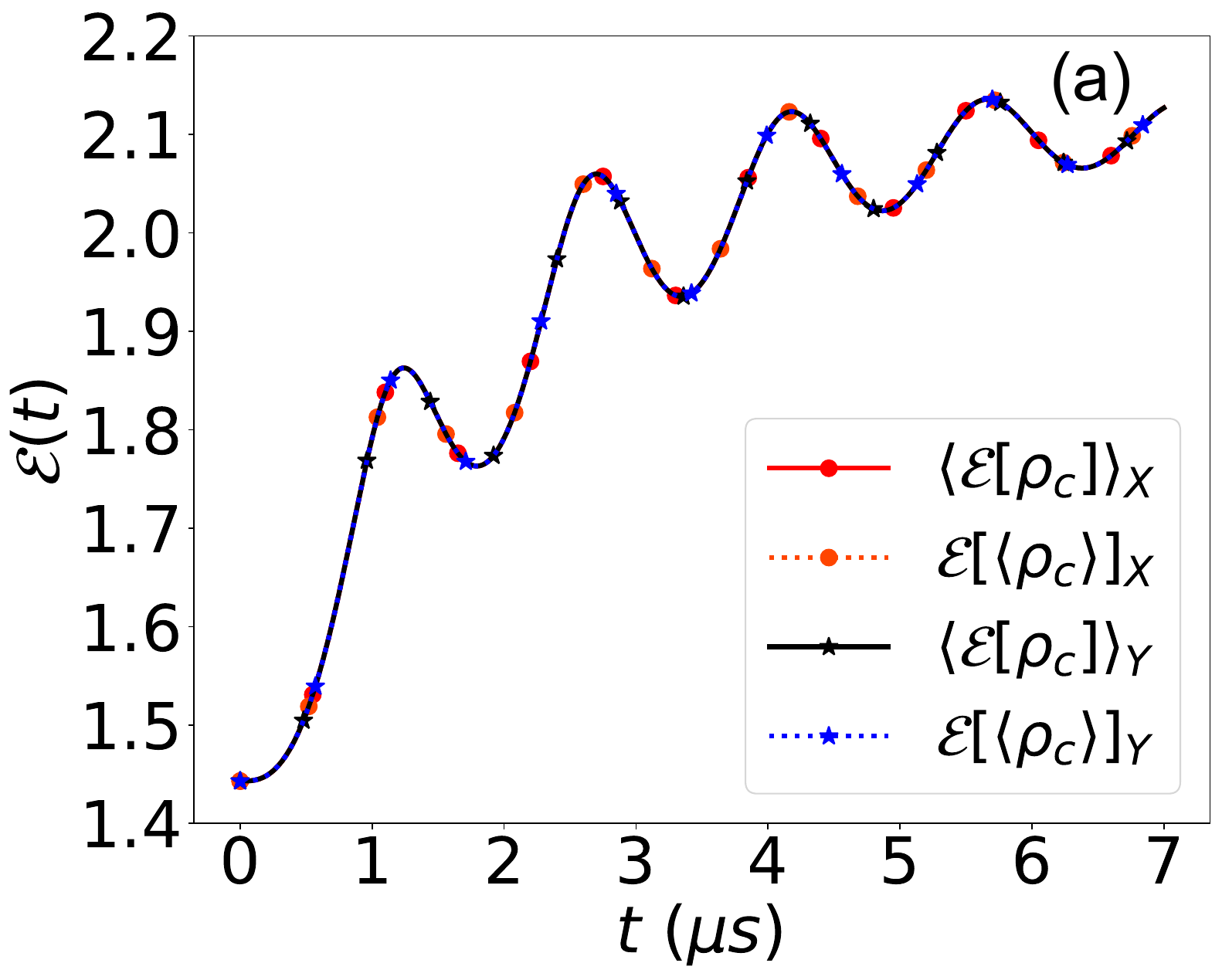}
    \includegraphics[width=0.48\linewidth]{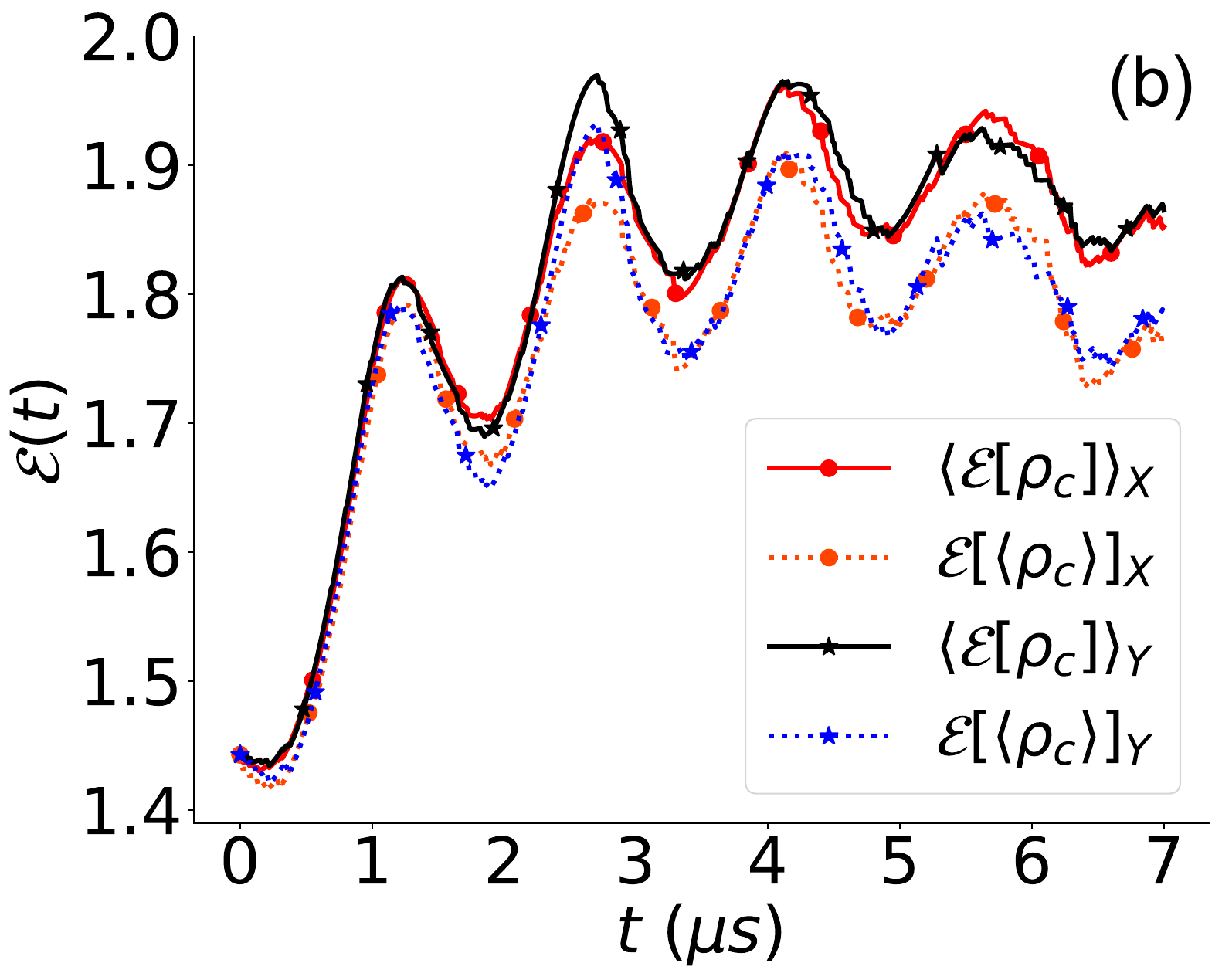}
    \caption{Ergotropy as a function of time for (a) no-jump and (b) jump case in the lifted-EP regime with $\Delta =1.5~\text{MHz}$. Solid lines represents $\langle \mathcal{E}[\rho_c] \rangle$, whereas dotted lines corresponds to $\mathcal{E}[\langle \rho_c \rangle]$. Red/orange and black/blue lines corresponds to $\sigma_x$ and $\sigma_y$-drive, respectively. Other parameters are $\Omega = 2~\text{MHz},~\Gamma_g =2~\text{MHz},~\Gamma_e=0.2~\text{MHz}$.}
    \label{fig:jump_ergo}
\end{figure}
\section{Quantum Jump Measurement Model}
\label{App:jump_measurement}
For efficient continuous quantum jump measurement case on three level system Fig.~\ref{fig: model} of main text, the general stochastic master equation is given as (See Ref.~\cite{role_of_inefficient} for details) 

\begin{figure}
    \centering
    \includegraphics[width=0.46\linewidth]{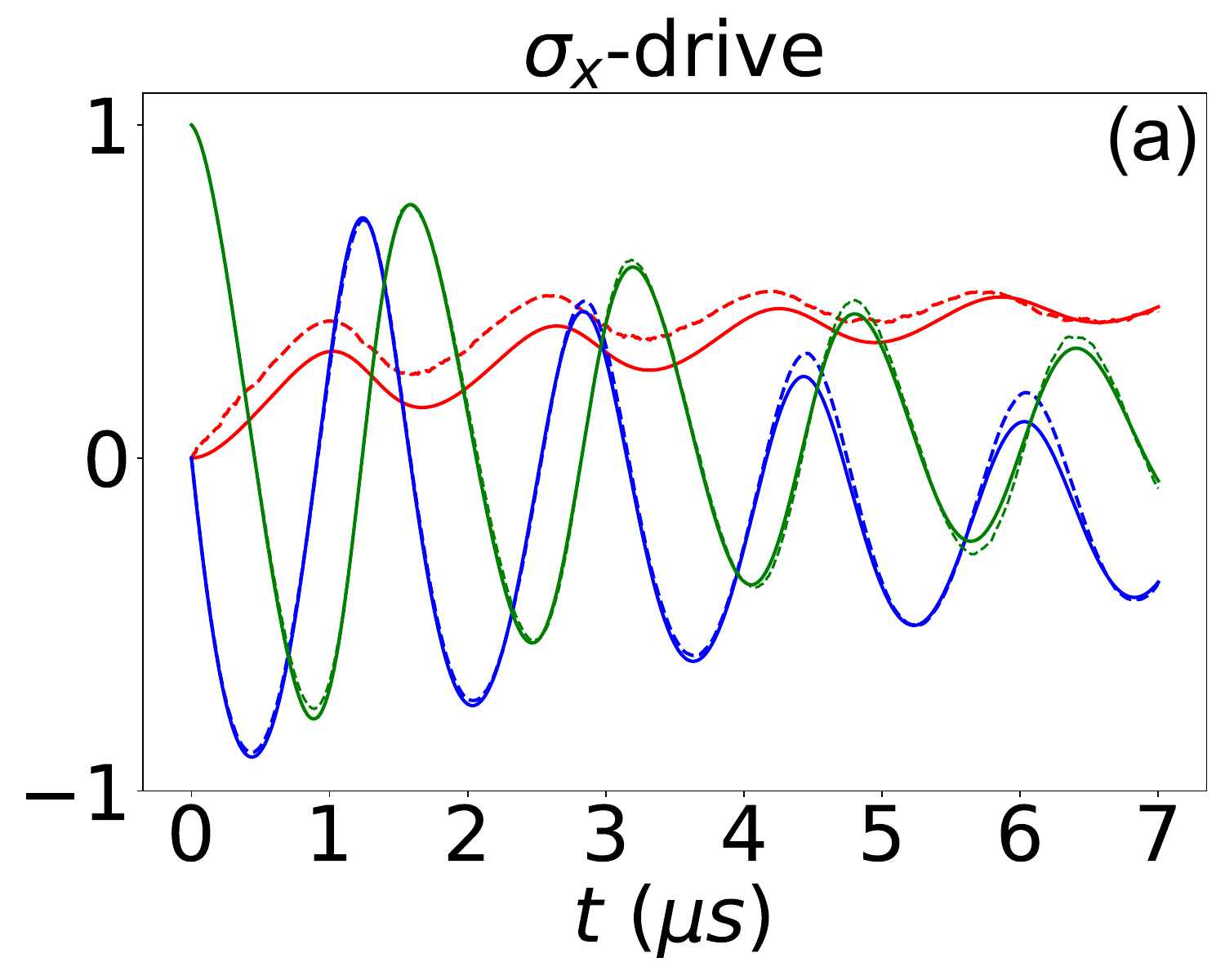}
    \includegraphics[width=0.46\linewidth]{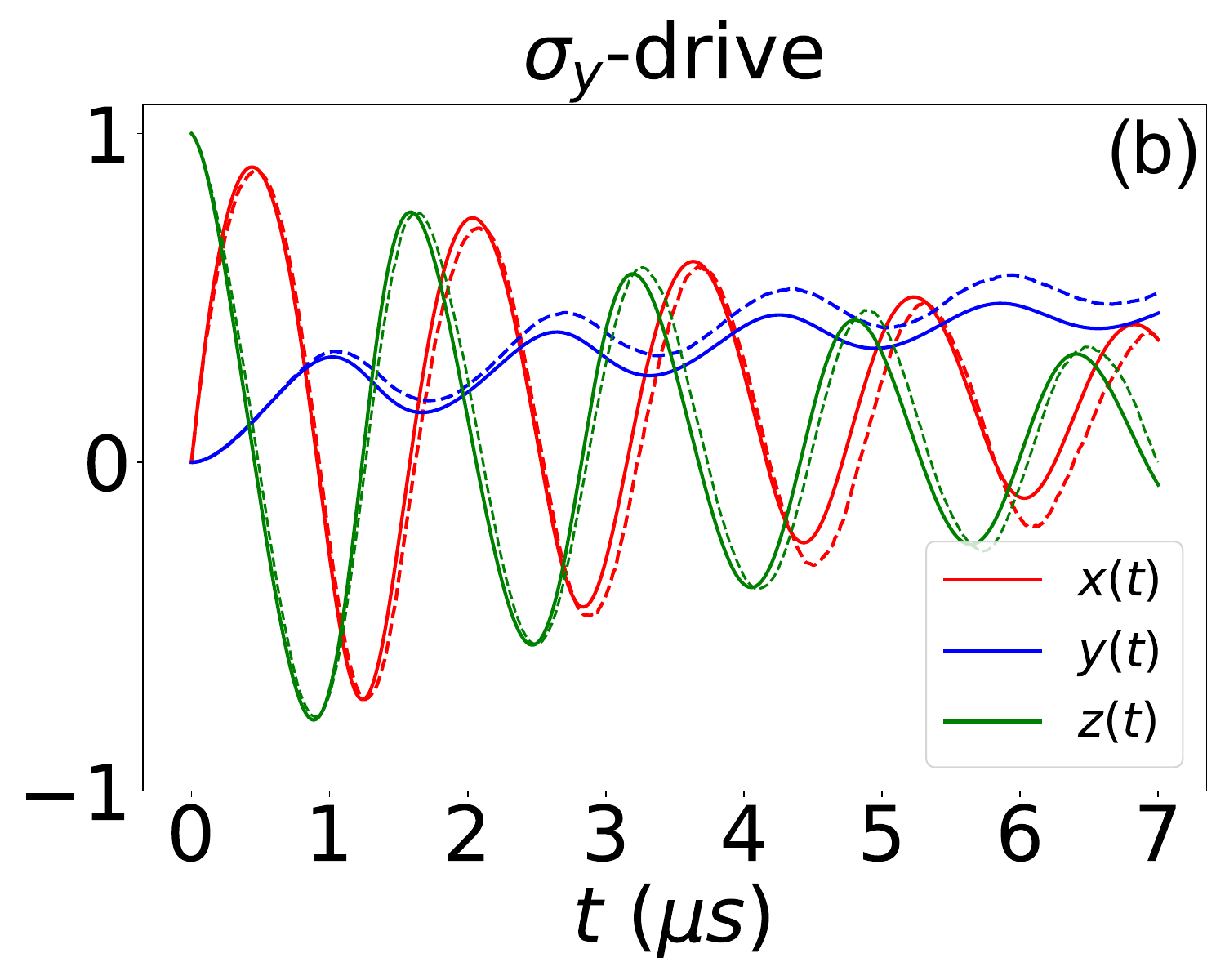}
    \includegraphics[width=0.48\linewidth]{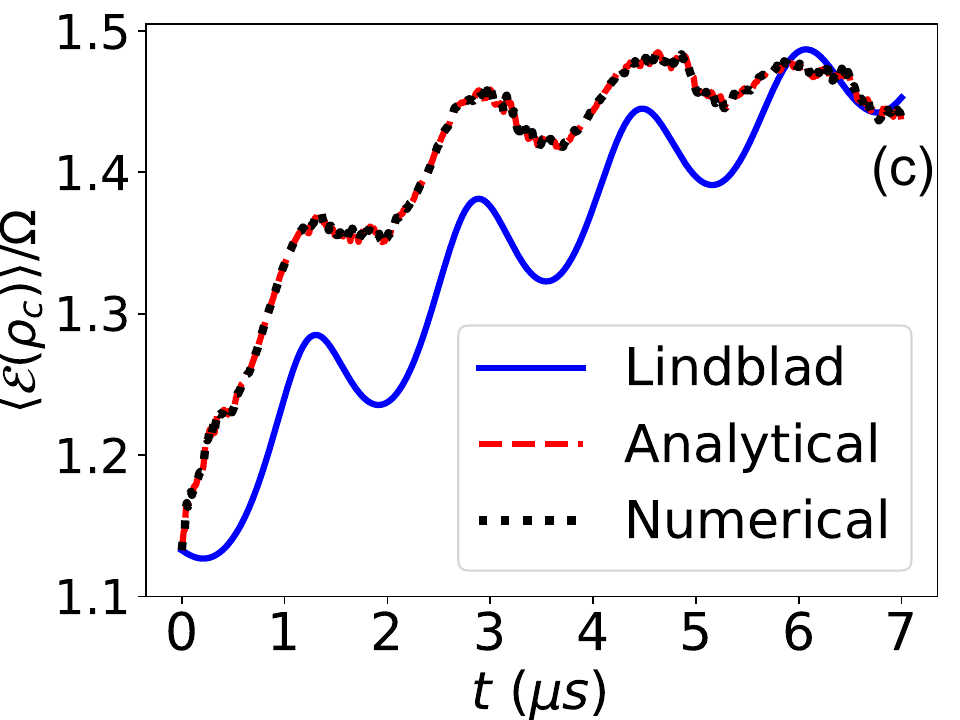}
    \includegraphics[width=0.48\linewidth]{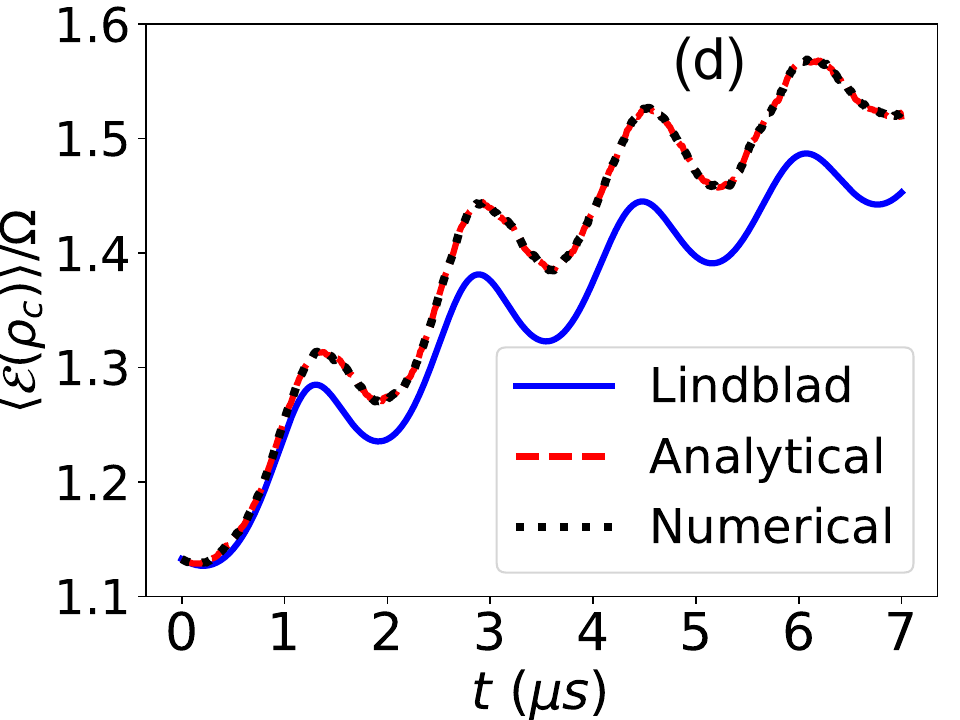}  
    \includegraphics[width=0.48\linewidth]{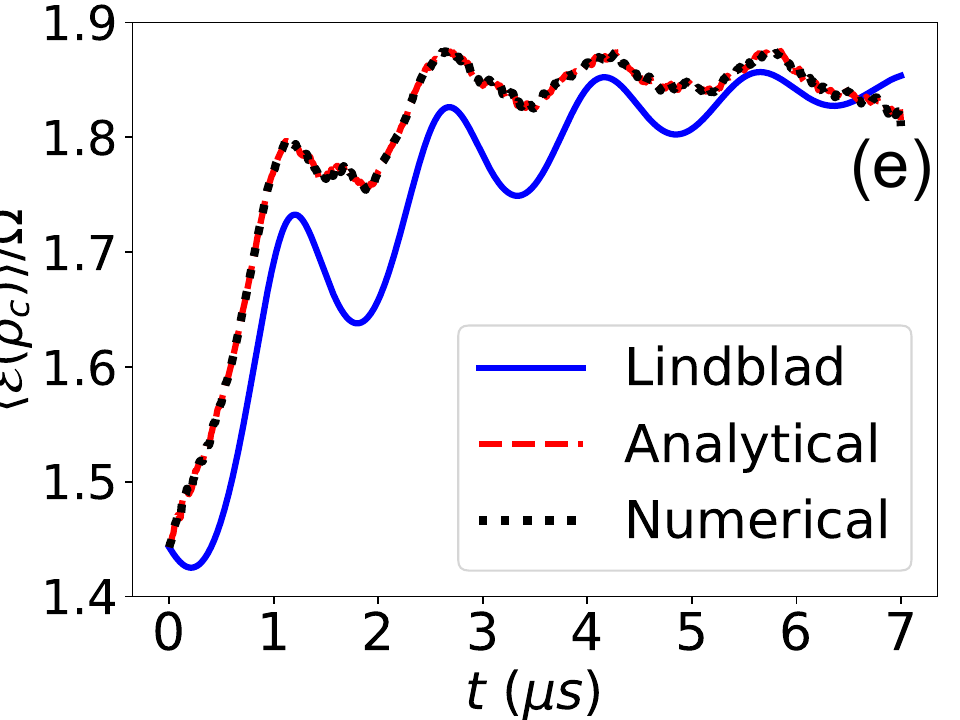}
    \includegraphics[width=0.48\linewidth]{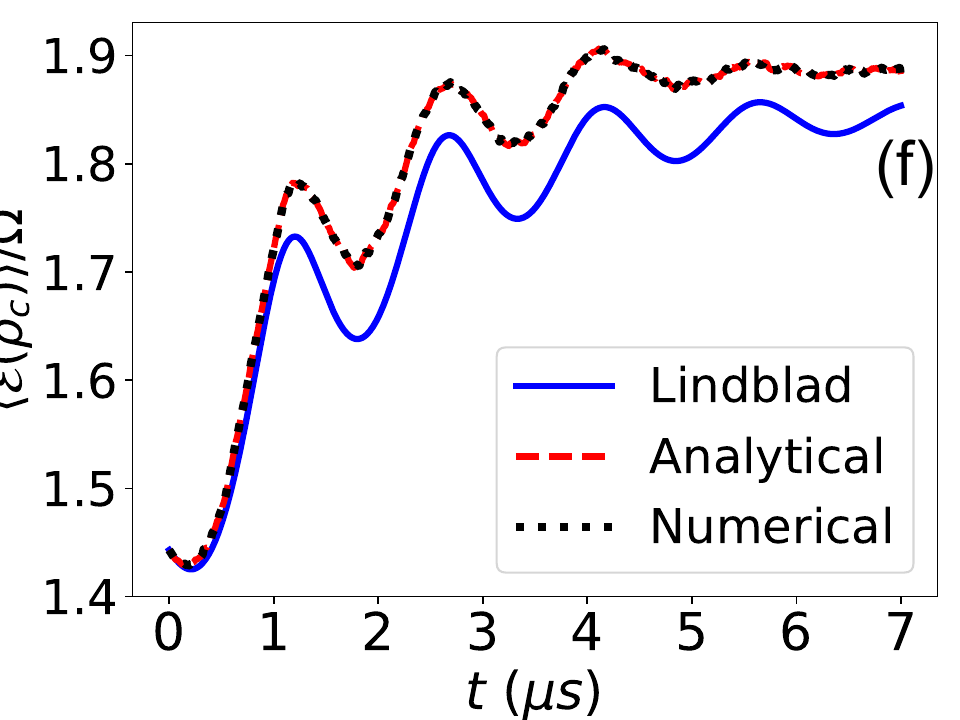}   
    \caption{Comparison of Lindblad and trajectory-based quantum battery dynamics. The left and right columns correspond to the $\sigma_x$- and $\sigma_y$-drives, respectively. Panels (a)-(b) and, (c)-(f) show the dynamics of the Bloch components and the ensemble-averaged ergotropy $\langle\mathcal{E}(\rho_c)\rangle$, respectively. In (a) and (b) solid lines corresponds to Lindblad dynamics whereas dashed lines represents the trajectory average. In panels (c)-(f), the blue solid, red dashed, and black dotted lines represent the Lindblad, analytical, and numerical trajectory dynamics, respectively. (a)-(d) corresponds to $\Delta=0.5~\text{MHz}$ case. whereas, (e) and (f) represents $\Delta=1.5~\text{MHz}$ case. Other parameters are $\Omega=2~\text{MHz},~\Delta=0.5~\text{MHz},~\Gamma_g =2~\text{MHz}$ and $\Gamma_e =0.2~\text{MHz}$.}
    \label{fig:lindblad_compare}
\end{figure}

\begin{multline}
d\rho_c = \Bigg( -i[H,\rho_c] -\frac{\Gamma_g}{2} \left\{\ket{1}\bra{1},\rho_c\right\} +\Gamma_g
            \left\langle\ket{1}\bra{1}\right\rangle\rho_c
\\
-\frac{\Gamma_e}{2} \left\{\ket{2}\bra{2},\rho_c\right\} +\Gamma_e \left\langle\ket{2}\bra{2}\right\rangle\rho_c\Bigg)dt
\\
+\left(\frac{\ket{0}\bra{1}\rho_c\ket{1}\bra{0}}{\left\langle\ket{1}\bra{1}\right\rangle}-\rho_c\right)dN_g
\\
+\left(\frac{\ket{1}\bra{2}\rho_c\ket{2}\bra{1}}{\left\langle\ket{2}\bra{2}\right\rangle}-\rho_c\right)dN_e .
\end{multline}

where $dN_g= \Gamma_g \langle \ket{1}\bra{1}\rangle dt$ and $dN_g= \Gamma_e \langle \ket{2}\bra{2}\rangle dt$ are the counting process for detector $D_1$ and $D_2$, respectively, in the measurement setup in Fig.~\ref{fig: model}  which are equal to $1$ when a photon is detected and $0$ otherwise. Solving this stochastic master equation numerically, one can post select trajectories with no $\ket{1}\rightarrow\ket{0}$ and $\ket{2}\rightarrow\ket{1}$ jumps. We name them no-jump trajectories. Furthermore, one can post-select only those trajectories with $\ket{2}\rightarrow\ket{1}$ jumps but excludes $\ket{1}\rightarrow \ket{0}$ jump trajectories. We name such trajectories as the jump trajectories.     
In Fig.~\ref{fig:jump_ergo}, ergotropy for the no-jump and jump case is ploted for lifted-EP case with $\Delta =1.5~\text{MHz}$. For no-jump case, the ergotropy is significantly larger than that of the jump case, due to the presence of decoherence. Furthermore, for the no-jump case, $\langle \mathcal{E}[\rho_c]\rangle= \mathcal{E}[\langle \rho_c \rangle]$ since the covariance between populations is exactly zero \cite{role_of_inefficient}. However, for the jump case, on the other hand, Jansen's inequality is satisfied. Note that the jump case is equivalent to the post-selected homodyne measurement case presented in Fig.~\ref{fig:4} and both dynamics match perfectly for the large number of post-selected trajectories.

\section{Comparison Between Lindblad and Trajectory Dynamics}
\label{app:compare_lindblad}

The Lindblad dynamics of the effective non-Hermitian system is governed by Eq.~(\ref{eq:Lindblad_ME}), or equivalently by Eqs.~(\ref{eq:trajectory_eq_xdrive}) and (\ref{eq:trajectory_eq_ydrive}) for the $\sigma_x$- and $\sigma_y$-drives, respectively, upon taking $\langle dW\rangle=0$. In general, the ensemble average of the conditioned trajectories deviates from the corresponding Lindblad dynamics due to the nonlinear terms arising from the normalization of the conditional state in the SME, Eq.~(\ref{eq: SME}) \cite{homodyne_roson,role_of_inefficient}. In Fig.~\ref{fig:lindblad_compare}(a) and (b), we compare the Lindblad dynamics of the Bloch components with the ensemble-averaged trajectory dynamics. The observed deviation between the two descriptions arises from the finite variance of $z$ and the covariances between the Bloch-vector components induced by post-selection and normalization. The resulting deviation in the Bloch components also leads to a deviation between the mean ergotropy calculated from the Lindblad dynamics and that obtained from the analytical or numerical dynamics, as shown in Fig.~\ref{fig:lindblad_compare}(c)-(d). Note that the mean ergotropy is a linear function of the Bloch-vector components (since $|\zeta|=1$ for each pure-state trajectory) and, therefore, can be obtained by replacing the trajectory-dependent quantities $x$, $y$, and $z$ with their corresponding ensemble averages. Thus, the mean ergotropy can be directly compared with the ergotropy calculated from the Liouvillian dynamics. On the other hand, in Fig.~\ref{fig:lindblad_compare} (e) and (f), we compare the Lindblad and ensemble-averaged trajectory dynamics for $\Delta=1.5~\text{MHz}$ for \(\sigma_x\) and \(\sigma_y\) drive, respectively. It can be observed that, for the larger detuning, the deviation between the two dynamics is significantly reduced compared with the $\Delta=0.5~\text{MHz}$ case as discussed in Sec.~\ref{sec:results} (C) of the main text.



\bibliography{Ref}
\end{document}